\global\def\draftcontrol{0}

   \def\versionno{NP large N}

\catcode`\@=11

\expandafter\ifx\csname draftcontrol\endcsname\relax\global\def\draftcontrol{0} 
\fi 

{\count255=\time\divide\count255 by 60 
\xdef\hourmin{\number\count255} 
\multiply\count255 by-60\advance\count255 by\time 
\xdef\hourmin{\hourmin:\ifnum\count255<10 0\fi\the\count255}} 
\def\draftdate{\number\month/\number\day/\number\year\ \ \ \hourmin } 

\newcommand\makepapertitle{\par

  \begingroup 
    \renewcommand\thefootnote{\@fnsymbol\c@footnote}%
    \def\@makefnmark{\rlap{\@textsuperscript{\normalfont\@thefnmark}}}%
    \long\def\@makefntext##1{\parindent 1em\noindent 
            \hb@xt@1.8em{%
                \hss\@textsuperscript{\normalfont\@thefnmark}}##1}%
     \newpage 
     \global\@topnum\z@   
     \@makepapertitle 
     \thispagestyle{empty}\@thanks 
  \endgroup 
  \setcounter{footnote}{0}%
  \global\let\thanks\relax 
  \global\let\makepapertitle\relax 
  \global\let\@makepapertitle\relax 
  \global\let\@thanks\@empty 
  \global\let\@author\@empty 
  \global\let\@date\@empty 
  \global\let\@title\@empty 
  \global\let\title\relax 
  \global\let\author\relax 
  \global\let\date\relax 
  \global\let\and\relax 
  \def\version{\let\version\@version\@gobble} 
} 
\def\@makepapertitle{%
  \newpage 
   \ifnum\draftcontrol=1 {} 
   \version\versionno 
   \vskip 5.5em%
   \else 
   \hfill\hbox to 3cm {\parbox{4.5cm}{\@pubnum}\hss}%
   \vskip 6.5em%
   \fi 
   \begin{center}%
   \let \footnote \thanks 
      {\hskip -0\textwidth \hbox to 1\textwidth%
        {\centerline{\Large\bf{\noindent\@title}}}}%
     \vskip 2em%
     {\normalsize
       \lineskip .5em%
       \begin{tabular}[t]{c}%
         \@author 
       \end{tabular}\par}%
     \vskip 1.5em%
     {\@bstract}%
     \end{center}%
     \vfill
     \@date%
     \vskip 1.5em%
   \par 
} 

\gdef\@pubnum{} 
\def\pubnum#1{%
  \gdef\@pubnum{#1}} 

\gdef\@bstract{} 
\def\Abstract#1{%
  \gdef\@bstract{%
   \parbox{\textwidth-0pc}{%
   \centerline{\bf Abstract}\penalty1000 
   \noindent
   \renewcommand\baselinestretch{1.0} 
   {#1}}} 
} 

\gdef\@email{}
\def\email#1{%
   \gdef\@email{%
   Email: {\tt #1}}
}

\def\ps@paper{\let\@mkboth\@gobbletwo%
     \ifnum\draftcontrol=1 
        \def\@oddfoot{\hbox to \textwidth{\tiny \versionno \hfil\tiny\draftdate}%
        \hskip -\textwidth \hbox to \textwidth{\hfil\rm\thepage\hfil}}%
     \else\def\@oddfoot{\hbox to \textwidth{\hfil\rm\thepage\hfil}} 
     \fi 
     \let\@evenfoot\@oddfoot 
} 

\def\body{\clearpage 
          \pagestyle{paper} 
        } 
\newenvironment{acknowledgments}{%
\vskip 3.25ex 
\addcontentsline{toc}{section}{Acknowledgments}
\noindent {\bf Acknowledgments} 
} 

\def\@version#1{\ifnum\draftcontrol=1 
\typeout{}\typeout{#1}\typeout{} 
\vskip3mm\centerline{\hbox{\fbox{\normalsize{\tt DRAFT -- #1 -- } 
                   {\draftdate}}}}\vskip3mm 
\fi} 
\let\version\@version 
\long\def\eqlabel#1{\ifnum\draftcontrol=1 
                    \tag@false  
                    \tag*{(\theequation) \hbox to -0.2cm{\hspace{0cm}\small{#1}\hss}} 
                    \refstepcounter{equation}  
                    \edef\@currentlabel{\theequation} 
                    \ltx@label{#1}          
                    \else 
                    \label{#1} 
                    \fi 
                    } 
\let\st@bibitem\@bibitem 
\let\st@lbibitem\@lbibitem 
\ifnum\draftcontrol=1 
  \def\@bibitem#1{%
    \st@bibitem{#1}\a@@label{#1}\ignorespaces} 
  \def\@lbibitem[#1]#2{%
    \st@lbibitem[#1]{#2}\a@@label{#2}\ignorespaces} 
  \def\a@@label#1{%
    \gdef\a@lab{\smash{\normalfont\small#1}} 
    \ifvmode 
      \if@inlabel 
        \global\setbox\@labels\hbox{%
          \llap{\a@lab\let\a@lab\relax 
                \kern\@totalleftmargin\kern\marginparsep}%
          \box\@labels}%
      \fi 
    \fi} 
\fi 

\documentclass[12pt,letterpaper]{article} 

\usepackage{amsmath,bm,amsfonts,amssymb,array,calc,amsthm,rotating,amscd}
\usepackage{epsfig,psfrag} 
\usepackage{rotating}
\usepackage{amscd}
\usepackage{graphicx}
\usepackage{color}
\usepackage[colorlinks=false]{hyperref}

\renewcommand\baselinestretch{1.25} 
\renewcommand\section{\@startsection {section}{1}{\z@}%
                                   {-3.5ex \@plus -1ex \@minus -.2ex}%
                                   {2.3ex \@plus.2ex}%
                                   {\normalfont\large\bfseries}} 
\renewcommand\subsection{\@startsection{subsection}{2}{\z@}%
                                   {-3.25ex\@plus -1ex \@minus -.2ex}%
                                   {1.5ex \@plus .2ex}%
                                   {\normalfont\normalsize\bfseries}} 
\renewcommand\subsubsection{\@startsection{subsubsection}{3}{\z@}%
                                   {-3.25ex\@plus -1ex \@minus -.2ex}%
                                   {1.5ex \@plus .2ex}%
                                   {\normalfont\normalsize\it}} 
\renewcommand\paragraph{\@startsection{paragraph}{4}{\z@}%
                                   {-1.75ex\@plus -1ex \@minus -.2ex}%
                                   {1ex \@plus .2ex}%
                                   {\normalfont\normalsize\bf}} 
\renewcommand\subparagraph{\@startsection{subparagraph}{5}{\z@}%
                                   {-1.25ex\@plus -0ex \@minus -.2ex}%
                                   {-2ex \@plus .2ex}%
                                   {\normalfont\normalsize\it}}

\numberwithin{equation}{section}

\long\def\@makecaption#1#2{%
  \vskip\abovecaptionskip
  \sbox\@tempboxa{{\bf #1:} #2}%
  \ifdim \wd\@tempboxa >\hsize
    {\small\bf #1:} {\small #2}\par
  \else
    \global \@minipagefalse
    \hb@xt@\hsize{\hfil\box\@tempboxa\hfil}%
  \fi
  \vskip\belowcaptionskip}

\renewcommand*\l@section[2]{%
  \ifnum \c@tocdepth >\z@
    \addpenalty\@secpenalty
    \addvspace{.5em \@plus\p@}%
    \setlength\@tempdima{1.5em}%
    \begingroup
      \parindent \z@ \rightskip \@pnumwidth
      \parfillskip -\@pnumwidth
      \leavevmode \bfseries
      \advance\leftskip\@tempdima
      \hskip -\leftskip
      #1\nobreak\hfil \nobreak\hb@xt@\@pnumwidth{\hss #2}\par
    \endgroup
  \fi}
\renewcommand*\l@subsection{\addvspace{.0em \@plus\p@}\@dottedtocline{2}{1.5em}{2.3em}}
\renewcommand*\l@subsubsection{\addvspace{-.2em \@plus\p@}\@dottedtocline{3}{3.8em}{3.2em}}

\def\hepth#1{\href{http://xxx.arxiv.org/abs/hep-th/#1}{{arXiv:hep-th/#1}}}

\def\math#1{\href{http://xxx.arxiv.org/abs/math/#1}{{arXiv:math/#1}}}

\def\arxiv#1#2{\href{http://xxx.arxiv.org/abs/#1}{{arXiv:#1 [#2]}}}

\definecolor{refcol}{rgb}{0.2,0.2,0.8}
\definecolor{eqcol}{rgb}{.6,0,0}
\definecolor{purple}{cmyk}{0,1,0,0}

\gdef\@citecolor{refcol}
\gdef\@linkcolor{eqcol}
\def\colorlinkspurple{\gdef\@urlcolor{purple}}
\def\colorlinksblue{\gdef\@urlcolor{blue}}
\def\colorlinksred{\gdef\@urlcolor{red}}

\def\ie{{\it i.e.}} 
\def\eg{{\it e.g.}} 
\def\etc{{\it etc.}}
\def\cf{{\it cf.}}

\def\revise#1       {\raisebox{-0em}{\rule{3pt}{1em}}%
                     \marginpar{\raisebox{.5em}{\vrule width3pt\ 
                     \vrule width0pt height 0pt depth0.5em 
                     \hbox to 0cm{\hspace{0cm}{%
                     \parbox[t]{4em}{\raggedright\footnotesize{#1}}}\hss}}}}

\def\ii           {{\it i}}

\def\Re           {{\rm Re\hskip0.1em}} 
\def\Im           {{\rm Im\hskip0.1em}}

\def\sqr#1#2{{\vcenter{\vbox{\hrule height.#2pt   
 \hbox{\vrule width.#2pt height#1pt \kern#1pt 
 \vrule width.#2pt}\hrule height.#2pt}}}}

\newcommand{\R}{\mathbb R}
\newcommand{\Z}{\mathbb Z}

\newcommand{\N}{\mathbb N}
\renewcommand{\S}{\mathbb S}

\newcommand{\Q}{\mathbb Q}

\newcommand{\Ical}{\mathcal I}

\newcommand{\Fcal}{\mathcal F}
\newcommand{\Tcal}{\mathcal T}

\newcommand{\Wcal}{\mathcal W}

\newcommand{\Ncal}{\mathcal N}

\newcommand{\ep}{\epsilon}

\newcommand{\ul}{\underline}

\newcommand{\beq}{\begin{equation}}
\newcommand{\eq}{\end{equation}}
\newcommand{\req}[1]{(\ref{#1})}

\catcode`\@=12 

\begin{document} 


\title{Exact Chern-Simons / Topological String duality}

\pubnum{
SNUTP15-004
}
\date{June 2015}

\author{
Daniel Krefl$^a$ and Ruben L. Mkrtchyan$^b$\\[0.2cm]
\it  $^a$ Center for Theoretical Physics, SNU, Seoul, South Korea\\
\it  $^b$ Yerevan Physics Institute, Yerevan, Armenia\\
}

\Abstract{
We invoke universal Chern-Simons theory to analytically calculate the exact free energy of the refined topological string on the resolved conifold. In the unrefined limit we reproduce non-perturbative corrections for the resolved conifold found elsewhere in the literature, thereby providing strong evidence that the Chern-Simons / topological string duality is exact, and in particular holds at arbitrary $N$ as well. In the refined case, the non-perturbative corrections we find are novel and appear to be non-trivial. We show that non-perturbatively special treatment is needed for rational valued deformation parameter. Above results are also extend to refined Chern-Simons with orthogonal groups.    
}

\makepapertitle

\body

\version\versionno

\vskip 1em



\section{Introduction}
By now, the large $N$ duality between $SU(N)$ Chern-Simons on $\S^3$ and the (perturbative) topological string on the resolved conifold of \cite{GV98} is well established. Over the years this duality has been extended in several ways, like for example to $SO/Sp$ gauge groups \cite{SV00}, leading at large $N$ to orientifolds of the topological string.

More recently, triggered by work of Nekrasov on $\Ncal=2$ supersymmetric gauge theory \cite{N02}, it became clear that there should exist a sort of refined topological string, being a one-parameter deformation of the usual topological string \cite{HIV03,IKV07} (for a brief exposition, see \cite{KW14}). The deformation parameter is usually denoted as $\beta$. In turn, this led Aganagic-Shakirov \cite{AS11,AS12a} to propose a refined Chern-Simons theory yielding at large $N$ the free energy of the refined topological string on the resolved conifold. Of course, at $\beta=1$ the original large $N$ dualities are recovered. In fact, the refined Chern-Simons theory can be defined for all $ADE$ groups \cite{AS12}. In particular, for $D_N$ this leads at large $N$ to a refinement of topological string orientifolds.

Perhaps less known, in a series of works a novel universal formulation of Chern-Simons theory on $\S^3$ has been put forward \cite{MV12, M13, M13b}. Here, universal refers to the fact that all the partition functions of Chern-Simons with arbitrary classical or exceptional simple gauge groups can be recovered from the universal Chern-Simons partition function under specialization of parameters. Quite surprisingly, the universal Chern-Simons theory does not only include the usual Chern-Simons theories, but, after some extension of range of parameters, as well the refined versions thereof, as shown in \cite{KS13}.

By construction, the universal Chern-Simons partition function constitutes an integral representation of the partition functions and thereby provides an analytic continuation in the parameters, \eg, simple Lie groups are now parametrized by the two-dimensional Vogel's plane \cite{V99}. This generalizes the old $N\rightarrow -N$ duality of gauge theories with orthogonal and symplectic groups \cite{M81}, and leads in particular to a suggestion of an extension of gauge/string duality to exceptional groups \cite{M14}. Furthermore, the integral representation is very well suited to study non-perturbative aspects of the large $N$ duality, see \cite{M13b, M14}.

In this work we continue this line of research by using the universal Chern-Simons integral representation to analytically calculate non-perturbative corrections to the refined Chern-Simons theory at large $N$, thereby proposing the non-perturbative completion of the refined topological string on the resolved conifold. However, one should keep in mind that for integer $N$ the non-perturbative corrections vanish, as the analytically continued partition function at integer $N$ coincides with the initial Chern-Simons partition function. At $\beta=1$ we can compare to recently renewed efforts to find the non-perturbative completion of the topological string, in particular to \cite{HMMO13,H15}. However, we like to stress that our approach is entirely analytic and does not, in contrast to other works in the literature, rely on any subtle combination of quantization, approximation and numerics. In fact, we do not even perform a genus (or more generally trans-series) nor large $N$ expansion, but directly recover for $SU(N)$ gauge group the refined Gopakumar-Vafa expansion of the resolved conifold (similar as previously in \cite{M13b, M14} for the unrefined case) and, most importantly, a non-perturbative completion thereof, via simple residue calculations. In particular, one may see this as the simplest example of a true (\ie, independent of $N$) gauge / string duality. Similar results are obtained for $SO(N)$ gauge group, leading to the prediction of the non-perturbative completion of orientifolds of the refined topological string on the resolved conifold.

The outline is as follows. In the next section we will recall the basic definitions of universal Chern-Simons theory. In particular, we will rewrite the known integral representation in a more convenient form, see section \ref{Generalities}, which will allow us to express the integral representation directly as a sum of residue, or, alternatively in terms of multiple sine functions, as discussed in section \ref{msineSec}. In the following two sections examples are discussed. Namely, in section \ref{ANsec} we calculate the universal partition function for unitary groups as a sum of residue, yielding the refined Gopakumar-Vafa expansion plus non-perturbative terms, with respect to the string coupling parameter. It is here where we establish a remarkable exact coincidence with the conjectured non-perturbative completion of the topological string on the resolved conifold. In Section \ref{Dn} we extend the results to orthogonal groups. Section \ref{QL} is devoted to the quantum limit of \cite{K13} applied to refined Chern-Simons, corresponding at large $N$ to the well-known Nekrasov-Shatashvilli limit. In particular a S-dual like relation between the perturbative and non-perturbative parts of the free energies will be discussed. Some more general remarks will also be given in this section. In appendixes we present some series identities and supplemental details about multiple gamma and sine functions, used heavily in the body of the paper.

\section{Universal Chern-Simons}
\subsection{Generalities}
\label{Generalities}
Recall that the universal Chern-Simons free energy $\Fcal$ reads \cite{MV12}
\beq\eqlabel{uCSF}
\Fcal=\Fcal^I-\Fcal^{II}-\frac{d}{2} \log \left(t/\delta\right)\,,
\eq
with 
$$
\Fcal^I:=\int_0^\infty \frac{dx}{x(e^x-1)} \left(f(x/\delta)-d\right)\,,\,\,\,\,\,\Fcal^{II}:=\int_0^\infty \frac{dx}{x(e^x-1)} \left( f(x/t)-d\right)\,,
$$
and
\beq
\begin{split}
f(x)&:=\frac{\sinh\left(\frac{x(v_1-2t)}{4}\right)\sinh\left(\frac{x(v_2-2t)}{4}\right)\sinh\left(\frac{x(v_3-2t)}{4}\right)}{\sinh\left(\frac{x v_1}{4}\right)\sinh\left(\frac{x v_2}{4}\right)\sinh\left(\frac{x v_3}{4}\right)}\,,\\
d&:=\frac{(v_1-2t)(v_2-2t)(v_3-2t)}{v_1 v_2 v_3}\,,
\end{split}
\eq
where we defined the effective coupling constant $\delta:=\kappa+t$ with $\kappa$ the usual Chern-Simons coupling constant. The parameters $v_i$ occurring in $f(x)$ and $d$ are referred to as Vogel's parameter. For particular choices of $v_i$ one can recover from \req{uCSF} the free energy of Chern-Simons theory on $\S^3$ with all simple Lie gauge groups. In particular, this requires to impose Vogel's condition $t=v_1+v_2+v_3$. Then $t$ is identified, in a special normalization,  with the dual Coxeter number $h$ of the corresponding simple Lie algebra. Normalization mentioned is called minimal one and is defined by the only negative Vogel parameter (usually $v_1$ below) to be equal to $-2$.  As discovered in \cite{KS13}, \req{uCSF} also includes the refined Chern-Simons theories of \cite{AS11,AS12} at appropriate values of parameters, though Vogel's condition will not be satisfied anymore in the refined case. 

It is convenient to rewrite $\Fcal$ as follows. We redefine $x\rightarrow t x/\delta$ in $\Fcal^{II}$ such that
$$
\Fcal^{II}=\int_0^\infty \frac{dx}{x(e^{t x/\delta}-1)} \left(f(x/\delta)-d\right)\,.
$$
Using the relation
\beq\eqlabel{cotmcotId}
\frac{1}{e^{b x}-1}-\frac{1}{e^{a x} -1} = \frac{e^{a x}-e^{b x}}{(e^{a x}-1)(e^{b x }-1)}=\frac{\sinh\left(\frac{x(a-b)}{2}\right)}{2\sinh\left(\frac{x a}{2}\right)\sinh\left(\frac{x b}{2}\right)}\,,
\eq
and making use of that the combined integrand is even under $x\rightarrow -x$, 
we can write $\Fcal$ as, 
\beq\eqlabel{FVpreFinal}
\Fcal=-\frac{d}{2} \log \left(t/\delta\right)+\frac{1}{4}\int_{\R_+} \frac{dx}{x} \frac{\sinh\left(x(t-\delta)\right)}{\sinh\left(x t \right)\sinh\left(x \delta\right)} \left(f(2x)-d\right)\,,
\eq
where we deformed the integration range to pass the origin on an infinitesimal semi-circle in the upper half of the complex plane. We will refer to the deformed contour as $\R_+$. (We could have equally well deformed to pass along in the lower half, which we will denote as $\R_-$). We further rescaled $x\rightarrow 2x \delta$. 

As we will show in section \ref{msineSec}, the integration over the $d_V$ term in fact cancels against the $\log$ term in \req{FVpreFinal}, such that we obtain the (novel) neat expression 
\beq\eqlabel{FintRep}
\boxed{
\Fcal=\frac{1}{4}\int_{\R_+} \frac{dx}{x} \frac{\sinh\left(x(t-\delta)\right)}{\sinh\left( x t \right)\sinh\left(x \delta\right)} f(2x)
}\,,
\eq
for the universal Chern-Simons free energy. One should note that the proof of equivalence between the integral representation \req{uCSF}, and so \req{FintRep}, and the ordinary (refined) Chern-Simons partition functions, performed in \cite{M13} and \cite{KS13}, is completely analytical and exact. Besides that it simultaneously encodes the usual and refined Chern-Simons theories in an unified way, other benefits of the integral representation \req{FintRep} are that it naturally extends Chern-Simons theory to non-integer and/or negative values of $N$ and the refinement parameter $\beta$, to non-integer values of $\delta$ (\cf, \cite{W10}), and more generally to wide ranges of complex values of parameters. Particularly  it allows an exact large $N$ expansion without any need to perform a semi-classical analysis. This, in particular the last point, will become more clear below, where we will discuss  ways to explicitly evaluate the integral occurring in \req{FintRep} as a sum of residue. 

Note first that it is not easy to establish convergence restrictions for the integral in \req{FintRep} for arbitrary values of parameters. Here, we make the assumption that $2t > v_i, \forall i$ and $t > 0$. In particular, this assumption is satisfied for the parameters corresponding to $SU(N)$ and $SO(N)$ (with $\beta > 0$). We further restrict to $\kappa > 0$ corresponding to $\delta > t > 0$ (we can also discuss $\kappa <0$ similarly, but omit it here). Denoting the integrand of \req{FintRep} as $\Ical$, we can estimate under the above restrictions on parameters that for large $x$ one has $\Ical \sim \frac{1}{x} e^{\frac{x}{2}(2t-\sum_{i}(v_i+|v_i|))}$. This leads us to the condition (for the real parts)
\beq\eqlabel{kpbound}
\boxed{
  \kappa >0\,:\,\,\,\,\, 2t- \sum_{i=1}^3(v_i+|v_i|) < 0
}\,,
\eq
needed for a convergent integral. For parameters of refined $SU(N)$ and $SO(N)$ this condition is indeed satisfied, \cf, \cite{KS13} and the example sections later on.

In order to show that one can directly rewrite the integral \req{FintRep} as a sum of residue, \ie, 
\beq
\eqlabel{FresFormula}
\boxed{
\Fcal = 2\pi\ii \sum_{\{x^*_+\}}{\rm Res}(\Ical, x^*_+)
}\,, 
\eq
where $\{x^*_+\}$ denotes the set of poles of the integrand, one has to establish that one can deform the integration contour without pickup up an extra contribution. That this is indeed the case follows from the multiple sine representation we will discuss below. One should however note that the pole structure of the integrand is subtle. Depending on the particular values the parameters take (\eg, integer multiplies of each other), enhancement to higher order poles may occur for some subsets of poles.

\subsection{Multiple sine representation}
\label{msineSec}
Though \req{FresFormula} looks at first glance very simple, the expressions resulting from taking residue directly of $\Ical$ are in fact not as easy as one might expect. In particular, one needs to invoke non-trivial summation identities to bring the result to canonical forms, used in literature. Following ideas of \cite{M13b}, it is however possible to find a closed simple expression for \req{FintRep} in terms of multiple sine functions for which the integral representation, and so the residue calculations, are somewhat different (but of course equivalent due to non-trivial identities), but already more or less in canonical form. The price to pay is the introduction of some mathematical machinery of multiple Barnes' gamma functions, multiple sine functions and their integral representations. This, however is justified by compactness of the resulting expressions, clear rules of transformations and mathematical rigorousness, in particular justifying that simple residue taking of $\Ical$ is indeed valid.

\paragraph{Definitions}

We need first to recall some basic definitions. Barnes' multiple $\zeta$-function \cite{B1904} is defined as
\beq\eqlabel{BzetaDef}
\zeta_r(z,s|\underline{w}) := \zeta_r(z,s|w_1,w_2,...,w_r)=
\sum_{n_1,...,n_r=0}^{\infty}\frac{1}{(z+w_1n_1+w_2n_2+\dots+w_r n_r)^s}\,.
\eq
Note that $\zeta_r(z,s|\underline{w})$ is well defined if all parameters $w_i$ lie on the same side of some straight line through the origin, and $\Re z >0,\, \Re s> r$. In similarity with the definition of Euler's gamma-function in terms of Riemann's zeta-function, one can use $\zeta_r(z,s|\underline{w})$ to define a multiple gamma-function
$$
\ln\Gamma_r(z|\ul w):=\partial_s\zeta_r(w,s|w_1,w_2,...,w_r)|_{s=0}\,.
$$
This definition follows \cite{R00} and differs from the original Barnes' one \cite{B1904} by some modular ``constant'', depending on parameters. 

Next, the multiple sine functions are defined via Barnes' gamma function as
\beq\eqlabel{mSineGdef}
S_r(z|\underline{\omega}):=\frac{\Gamma_r(|\omega|-z|\underline{\omega})^{(-1)^r}}{\Gamma_r(z|\underline{\omega})}\,,
\eq
where $|\underline{\omega}|:=\sum_{j=1}^{r}\omega_j$. Some useful identities for the multiple sine functions we will make us of in this section are
\beq
\eqlabel{SrIds}
\begin{split}
S_r(z|\ul w)&=S_r(|\ul w|-z|\ul w)^{(-1)^{r+1}}\,,\\
S_r(c z|c \ul w) &= S_r(z|\ul w)\,,\\
S_2(1|1,w)&=\sqrt{w}\,.
\end{split}
\eq 
(More identities are listed in \req{mSineAllIds}.)

According to \cite{NARU03,F12} the multiple sine functions $S_r$ possess for $r\geq 2$, $0< \Re w_j$, and $0<\Re z<|\ul w|$ an integral representation over the entire real line, bypassing the singularity at zero either in the upper or lower half of the complex plane, \ie,
\beq\eqlabel{SrIntRep}
\begin{split}
S_r(z|\underline{\omega})&=\exp\left( (-1)^r \frac{\pi i}{r!}B_{r,r}(z|\underline{\omega})+(-1)^r \int_{\R_+}\frac{dx}{x}\frac{e^{zx}}{\prod_{k=1}^r(e^{\omega_i x}-1)} \right)\\
&=\exp\left((-1)^{r-1} \frac{\pi i}{r!}B_{r,r}(z|\underline{\omega})+(-1)^r \int_{\R_-}\frac{dx}{x}\frac{e^{zx}}{\prod_{k=1}^r(e^{\omega_i x}-1)} \right)\,,
\end{split}
\eq
where $B_{r,r}$ refer to the generalized Bernoulli polynomials, defined via the generating function
\beq\eqlabel{GBernoulliDef}
\frac{x^{r}e^{zx}}{\prod_{j=1}^{r} (e^{w_jx}-1)}= \sum_{n=0}^{\infty}\frac{x^n}{n!}B_{r,n}(z|\ul w)\,.
\eq
Note that the Bernoulli terms occurring in \req{SrIntRep} are $\pm \ii\pi$ times the residue of the integrand at $x=0$, \ie, they are equal to half of the integral with same integrand over the small circle around $x=0$. These terms are necessary to have an equality between the two  different integral representation above. The mentioned integral representation of these Bernoulli terms leads to the similar to \req{SrIds}  identities for Bernoulli polynomials. In particular, we will need in the following discussion the identity \cite{NARU03}
\beq\eqlabel{BshiftId}
B_{r,n}(z| \ul w) = (-1)^{n} B_{r,n}(|\ul w| -z| \ul w)\,.
\eq
Using results of \cite{NARU03}, the logarithm of multiple sine functions can be expressed as a sum of residue, \ie,
\beq
\eqlabel{SrViaResidue}
\log S_r(z|\ul w) = (-1)^r \frac{\pi\ii}{r!} B_{r,r}(z|\ul w)+(-1)^r \sum_{\{x_*^+\}}{\rm Res}(\mathcal S(z|\ul w),x^+_*)\,.
\eq
According to \cite{NARU03}, this expression is valid provided $\Im z > \Im |\underline{w}|>0$. (Similar for the contour $\R_-$, for which one needs $0<\Im z < \Im|\ul w|$). However, there is one subtlety. Actually, the bound of \cite{NARU03} appears to be slightly too strong. Namely, we believe that one can relax the bound to $\geq$, via an improved estimation, performing for instance a similar discussion as in \cite{FV99} (see their section 3.5).

\paragraph{Quintuple sine free energy}

We now have everything at hand to rewrite \req{FintRep} in terms of multiple sine functions. Let us however first show that the $\log$ term in \req{FVpreFinal} indeed cancels against the integral over \req{cotmcotId}. We can rewrite (using the equalities in \req{cotmcotId})
$$
\frac{1}{4}\int_{\R_+} \frac{dx}{x} \frac{\sinh\left(x(t-\delta)\right)}{\sinh\left(x t\right)\sinh\left(x \delta\right)}= \frac{1}{2}\int_{\R_+} \frac{dx}{x}  \frac{e^{x t}}{(e^{t x}-1)(e^{\delta x }-1)}-\frac{1}{2}\int_{\R_+} \frac{dx}{x}  \frac{e^{\delta x}}{(e^{tx}-1)(e^{\delta x }-1)}\,.
$$
Note that we have due to the identity \req{BshiftId}
$$
B_{r,r}(t|t,\delta) - B_{r,r}(\delta|t,\delta) = (1-(-1)^{r}) B_{r,r}(\delta|t,\delta)\,,
$$
which equals zero for $r$ even. Hence, we can freely add this pair of Bernoulli polynomials to the expression above, thereby being able to rewrite
$$
\frac{1}{4}\int_{\R_+} \frac{dx}{x} \frac{\sinh\left(x(t-\delta)\right)}{\sinh\left(x t\right)\sinh\left(x \delta\right)}=\frac{1}{2}\log\left(\frac{S_2(t|t,\delta)}{S_2(\delta|t,\delta)} \right)=-\frac{1}{2}\log\left(\frac{t}{\delta}\right)\,,
$$
where we made use of the multiple sine integral representation \req{SrIntRep} and the identities given in \req{SrIds}. We conclude that the $\log$ term in \req{FVpreFinal} indeed cancels out. 

Finally, let us discuss how to express the universal Chern-Simons free energy \req{FintRep} entirely in terms of multiple sine functions. For that note that the integrand $\Ical$ is a product of five $\sinh$ factors. We can expand the $\sinh$ factors in terms of exponentials, yielding
$$
\Fcal = s(v)\sum_{\{\sigma\}} \,\frac{s(\sigma)}{2} \int_{\R_+} \frac{dx}{x} \frac{e^{(z_\sigma +\frac{1}{2}|\ul w|)\,x}}{\left(e^{2 x t}-1\right)\left(e^{2 x \delta }-1\right)\prod_{i=1}^3\left(e^{x |v_i|}-1\right)}\,,
$$
where $\{\sigma\}$ is the set of 16 tuples of length 4 of all possible sign combinations, \ie, $\sigma=\{\sigma_0,...,\sigma_3\}$ with $\sigma_i=\pm 1$, $s(\sigma):=\prod_{i=0}^3 \sigma_i$ the parity of the tuple ($s(v)$ is similarly defined),
\beq\eqlabel{zSigmaDef}
z_\sigma = \sigma_0(t-\delta)+\frac{1}{2}\sum_{i=1}^3\sigma_i (v_i-2t)\,,
\eq
and $|\ul w|=2(t+\delta)+\sum_{i=1}^3 |v_i|$\,. Now note that the set ${\sigma}$ has a $\Z_2$ symmetry identifying tuples differing by overall sign. Hence we can write
$$
\Fcal = s(v)\sum_{\{\sigma\}/ \Z_2} \,\frac{s(\sigma)}{2} \int_{\R_+} \frac{dx}{x} \frac{e^{(\frac{1}{2}|\ul w|+z_\sigma )\,x}+e^{(\frac{1}{2}|\ul w|-z_\sigma )\,x}}{\left(e^{2 x t}-1\right)\left(e^{2 x \delta }-1\right)\prod_{i=1}^3\left(e^{x |v_i|}-1\right)}\,.
$$
Clearly, the identities \req{SrIntRep} and \req{BshiftId} can be shifted by $z\rightarrow z+\frac{1}{2}|\ul w|$. In particular, we have
$$
B_{r,r}(|\ul w|/2 -z_\sigma |\ul w)+B_{r,r}(|\ul w|/2 +z_\sigma |\ul w)=(1+(-1)^r) B_{r,r}(|\ul w|/2 -z_\sigma |\ul w)\,,  
$$
vanishing for $r$ odd. Hence, we can again freely add such pairs of Bernoulli polynomials to the above summation. Making use of \req{SrIntRep} and \req{BshiftId}, we conclude
\beq\eqlabel{FVviaSine5}
\Fcal = s(v)\,  \log \prod_{\{\sigma\}/ \Z_2} \,S_5\left(z_\sigma+|\ul w|/2 \right|\left.2t,2\delta,|v_1|,|v_2|,|v_3|\right)^{s(\sigma) } \,.
\eq
Note that the condition $0<\Re|z|<|\ul w|$ needed to rewrite the integral representation in terms of multiple sine functions is equivalent to \req{kpbound}, as is evident from \req{zSigmaDef}.

Denoting as above the integrand in \req{SrIntRep} as $\mathcal S(z|\ul w)$, we finally arrive at
\beq\eqlabel{QSineRes}
\boxed{
\Fcal =  -s(v)\, \sum_{\{x^*_+\},\{\sigma\}/ \Z_2} s(\sigma) \left( {\rm Res}\left(\mathcal S(|\ul w|/2+z_\sigma|\ul w), x_+^*\right)+\frac{\ii \pi}{5!} B_{5,5}(|\ul w|/2+z_\sigma|\ul w) \right)
}\,,
\eq
where as before $\{x^*_+\}$ denotes the set of poles on the upper imaginary axis. Equation \req{QSineRes} constitutes the main result of this section. The expression \req{QSineRes} is simpler than \req{FresFormula}, because the numerator of the integrand $\mathcal S(z|\ul w)$ is not a product of trigonometric functions as in \req{FresFormula}, but just an exponential. In particular, due to the used multiple sine identities, the exponentials do not combine anymore to trigonometric functions. This implies the existence of non-trivial summation identities to map to the results obtained via the direct integration (\cf, appendix \ref{TrigSumIdsSec}).

\section{Example 1: $A_{N-1}$}
\label{ANsec}
The first example we consider is refined $A_{N-1}$, whose universal Chern-Simons representation has been derived in \cite{KS13}. Namely, the refined $A_{N-1}$ theory sits at 
$$
v_1=-2\,,\,\,\,\, v_2=2\beta\,,\,\,\,\,\, v_3=t=\beta N\,.
$$
In order to compare with topological string theory, it is convenient to introduce the parameter 
\beq\eqlabel{ANtHooft}
\mu:=\frac{t}{\delta}=\frac{\beta N}{\delta}\,.
\eq
Usually, this is the t'Hooft parameter kept fixed at large $N$. Since we do not perform a large $N$ expansion, for us $\mu$ is just a parameter. The integrand of \req{FresFormula} specializes under this choice of parameters to
$$
\Ical_A=\frac{\sinh\left(x\delta(\mu-1)\right)\sinh\left(x(\delta\mu+1)\right)\sinh\left(x(\delta\mu-\beta)\right)}{4x\sinh\left(\delta\mu x \right)\sinh\left(\delta x\right)\sinh\left(\beta x \right)\sinh\left(x\right)}\,.
$$
As for these choice of Vogel's parameters one pair of $sinh$ cancelled out, it is more convenient to write the corresponding partition function $Z:=e^{\Fcal}$ in terms of quadruple sine functions, rather than via the general expression \req{FVviaSine5} consisting of quintuple sines. Making similar considerations as in section \ref{msineSec}, we almost immediately deduce (after writing the $\sinh$ in terms of exponentials)
\begin{eqnarray}
Z_A^{-1}= S_4(\beta|\ul w)\,S_4(\delta|\ul w)\,S_4(1+\beta+\delta|\ul w)\,S_4(1+2\delta\mu|\ul w)\,,
\end{eqnarray}
with $\ul w=(1,\beta,\delta,\delta\mu)$.
Using recurrent relations and identities of multiple sine functions (\cf, appendix \ref{mgsf}),
the partition function can be simplified to
\beq\eqlabel{ZaS3final}
Z_A(\mu;\beta)=\frac{1}{\sqrt{\mu}}\frac{S_3(1+\delta\mu|1,\beta,\delta)}{S_3(\beta|1,\beta,\delta)}  \,.
\eq
Note that the multiple sine representation is convenient for investigating symmetries of the partition function. For example, at $\beta=1$ we clearly see the level-rank duality $k\leftrightarrow N$ (up to the prefactor). Another potentially interesting symmetry is the exchange $\beta \leftrightarrow \delta$, provided $t=\delta\mu=\beta N$ is invariant, \ie, if simultaneously $N\rightarrow N\beta/\delta$. 

The for us important symmetry of the partition function is with respect to the transformation
\beq\eqlabel{AnSymLongParas}
\begin{split}
\beta \rightarrow 1/\beta\,,\,\,\,\, N \rightarrow \beta N+1-\beta\,,\,\,\,\,\,  
\delta \rightarrow\delta / \beta\,,\,\,\,\,\,
\mu \rightarrow \mu + \frac{1-\beta}{\delta}\,.
\end{split}
\eq
Since this symmetry involves inversion of $\beta$, it turns via the relation between $\beta$ and the equivariant parameters of the $\Omega$-background \cite{DV09}
\beq\eqlabel{epsDef}
\ep_1=\sqrt{\beta}\, g_s\,,\,\,\,\,\,\ep_2=-\frac{1}{\sqrt{\beta}}\, g_s\,, 
\eq
into the $\Omega$-background symmetry $\ep_1 \leftrightarrow -\ep_2$.

Generally, at large $N$ the parameter $\mu$ is viewed as fixed t'Hooft coupling constant, \ie, it is assumed to not transform under \req{AnSymLongParas}. However, the $\beta$-inversion symmetry is still manifest. Namely, shifting and rescaling
\beq\eqlabel{AbetaSymmetrize}
\begin{split}
\mu&\rightarrow \bar\mu  -\frac{1}{2\delta}\left(1-\beta\right)\,,\\
\delta&\rightarrow \sqrt{\beta}\,\bar\delta\,,
\end{split}
\eq
we have in terms of fixed $\bar \mu$ and $\bar \delta$
\begin{eqnarray} 
Z_A(\bar\mu;\beta)=\frac{\beta}{\sqrt{\bar\mu-\frac{1}{2}(1-\beta)}} \frac{S_3(\frac{1}{2}(\sqrt{\beta}+\frac{1}{\sqrt{\beta}})+\bar\delta\bar\mu)|\frac{1}{\sqrt{\beta}},\sqrt{\beta},\bar\delta)}{S_3(\sqrt{\beta}|\frac{1}{\sqrt{\beta}},\sqrt{\beta},\bar\delta)}\,.
\end{eqnarray} 
Up to a logarithmic term originating from the first factor, we conclude that the exact free energy $\Fcal_A(\bar\mu;\beta)$ is invariant under
\beq\eqlabel{betaInversion}
\beta\rightarrow  \frac{1}{\beta}\,.
\eq 
Note that the transformation \req{AbetaSymmetrize} is the usual shift one has to perform to obtain an even powers of $g_s$ only expansion of refined free energies, \cf, \cite{KW10a,KW10b}. 

Let us continue to evaluate \req{ZaS3final}. For that, we make use of the integral representation \req{SrIntRep}, yielding
\beq\eqlabel{FAmSineInt}
\Fcal_A\sim \int_{\R^+}\frac{dx}{x}\frac{e^{x(1+\delta\mu)}-e^{x \beta}}{\left(e^{x}-1\right)\left(e^{x \beta}-1\right)\left(e^{x \delta}-1\right)}\,,
\eq
where we dropped for convenience the $\log\mu$ term and two generalized Bernoulli polynomials. We can solve the integral via summing over residue of the integrand, \cf, \req{SrViaResidue}. However, care has to be taken about what values the parameters take, as enhancement from simple poles to higher order poles may occur. In particular, for simplicity we will assume that $\delta\not\in \Q$. Note also, that as one ray of poles depend on $1/\delta\sim g_s$, there is a natural split into {\it perturbative} and {\it non-perturbative} poles, the latter being independent of $\delta$. Hence, there is a natural split as well of the free energy, \ie, 
$$
\Fcal_A(\mu;\beta)=\Fcal^P_A(\mu;\beta)+\Fcal^{NP}_A(\mu;\beta)\,.
$$
The residue calculation of \req{FAmSineInt} is straight-forward, except that one has to distinguish in the non-perturbative sector between $\beta$ rational or not. The former will introduce some additional calculational complications due to pole enhancement of some of the poles. (Nevertheless the free energy for rational $\beta$ can be derived as well by a non-trivial limit from the non-rational $\beta$ free energy.)

Instead of stating directly here the resulting expressions for $\Fcal_A^P$ and $\Fcal_A^{NP}$, it is instructive to calculate these free energies as well via the direct integration of $\Ical_A$ described in section \ref{Generalities}. Though this approach is on a technical level more complicated than using multiple sine functions, it is to some extent more illustrative, as for instance flop invariance is explicitly manifest, the Bernoulli polynomials are absent, and we can easily invoke trigonometric identities. 

As we assume $\delta\not\in \Q$, so we have $\mu\not\in \Q$. Equation \req{FresFormula} tells us that
$$
\Fcal_{A}=\,2\pi \ii\sum_{\{x^*_+\}}{\rm Res}(\Ical_A, x^*_+)\,.
$$
The set of poles consists of two rays of perturbative simple poles at $x^*_{p,1}=n\pi \ii/(\delta\mu)$ and $x^*_{p,2}=n\pi \ii/\delta$ (since $\mu\not\in\Z$) and two rays of non-perturbative poles at $x^{*}_{np,1}=n\pi \ii /\beta$ and $x^{*}_{np,2}=n\pi \ii$, where $n\in \N \setminus\{0\}$. Hence,
\beq
\Fcal^P_A(\mu;\beta):=R_{p,1}+R_{p,2}\,,\,\,\,\,\, \Fcal^{NP}_A(\mu;\beta):=R_{np,1}+R_{np,2}\,,
\eq
where $R$ denotes the residue sum of the respective ray of poles. (The sums here and below are actually understood as a limit of sum of partial sums of all objects involved, since infinities are cancelled in between them.) The perturbative contribution $\Fcal^P_A$ can be easily inferred, up to some technical subtlety. Namely, only the combined $R_{p,1}+R_{p,2}$ is finite over summation over the set of poles. In particular, taking partial sums implies that we pick up a left over contribution of $-\frac{1}{2}\log\mu$ in canceling the singularities against each other, which in fact exactly reproduces the prefactor of \req{ZaS3final}. Using trigonometric product-to-sum formula and addition identities, it is not hard to infer that we can write the combined residue summation as
\beq\eqlabel{FAPfinal}
\boxed{
\Fcal^P_A(\mu;\beta)=\log\frac{1}{\sqrt{\mu}}+\frac{1}{4}\sum_{n=1}^\infty\frac{\cos\left(n\pi(2\mu+(1-\beta)/\delta)\right)}{n \sin\left(\frac{n\pi}{\delta}\right)\sin\left(\frac{n\pi \beta}{\delta}\right)}-\frac{1}{4}\sum_{n=1}^\infty \frac{1+\cot\left(\frac{n\pi}{\delta}\right)\cot\left(\frac{n\pi \beta}{\delta}\right)}{n}
}\,.
\eq
Note that after shifting the free energy via \req{AbetaSymmetrize}, the $\beta$-inversion symmetry is clearly visible, and furthermore the free energy becomes explicitly flop invariant (\ie, invariant under $\bar\mu\rightarrow -\bar\mu$).

In order to make contact with the free energy of the refined topological string, we set 
$$
\bar\delta=\frac{2\pi\ii}{g_s}\,,
$$ 
and make use of the (non-trivial) sum identities \req{cosSumtoExpSumId} and \req{CotCotIdentity}, such that we arrive at
\beq\eqlabel{FAPconifold}
\Fcal_A^P(\bar\mu;\beta) \sim -\frac{1}{4}\sum_{n=1}^\infty \frac{Q^n}{n \sinh\left(\frac{n g_s}{2\sqrt{\beta}}\right)\sinh\left(\frac{n \sqrt{\beta} g_s}{2}\right)}+\frac{1}{2}\sum_{n=1}^\infty\frac{ e^{- n g_s (\sqrt{\beta}-1/\sqrt{\beta})/2}}{n \sinh\left(\frac{n g_s}{2\sqrt{\beta}}\right)\sinh\left(\frac{n \sqrt{\beta} g_s}{2}\right)}\,,
\eq
where we also defined $Q:=e^{-2\pi\ii \bar\mu}$ and dropped some terms consisting of generalized Bernoulli polynomials and as well the $\log\mu$ term. We recognize the refined Gopakumar-Vafa expansion and constant map contribution of the resolved conifold \cite{IKV07}.

Let us move on to the non-perturbative contribution. Note that we have to distinguish between $\beta\in \Q^*$ and $\beta\not \in \Q^*$, with $\Q^*:=\Q \setminus \{0\}$, as already mentioned above.

\paragraph{$\beta \not\in \Q$}
If $\beta$ is not a rational number, we just have two rays of non-perturbative simple poles at $x^*_{np,1}=n \pi \ii /\beta$ and $x^*_{np,2}=n \pi\ii$, leading to
\beq
\begin{split}
R_{np,1}&=-\frac{1}{2}\sum_{n=1}^\infty \frac{\sin\left(\frac{n\pi\delta(\mu-1)}{\beta}\right)\sin\left(\frac{n\pi\delta(\mu+1/\delta)}{\beta}\right)}{n\sin\left(\frac{n\pi}{\beta}\right)\sin\left(\frac{n\pi\delta}{\beta}\right)}\,,\\
R_{np,2}&=-\frac{1}{2}\sum_{n=1}^\infty \frac{\sin\left(n\pi \delta(\mu-1)\right)\sin\left(n\pi\delta(\mu-\beta/\delta)\right)}{n\sin\left(n\pi \beta\right)\sin\left(n\pi\delta\right)}\,.
\end{split}
\eq
Via invoking trigonometric identities, and under the redefinitions \req{AbetaSymmetrize}, we arrive at
\beq\eqlabel{FANPbnZfinal}
\boxed{
\begin{split}
\Fcal_A^{NP}(\bar \mu;\beta\not\in \Q)=&\,\frac{1}{4}\sum_{n=1}^\infty\left(\frac{\cos\left(\frac{n\pi\bar\delta(2\bar\mu-1)}{\sqrt{\beta}}+n\pi\right)}{n\sin\left(\frac{n\pi}{\beta}\right)\sin\left(\frac{n\pi\bar\delta}{\sqrt{\beta}}\right)}+\frac{\cos\left(n\pi \sqrt{\beta} \bar\delta(2\bar\mu-1)+n\pi\right)}{n\sin\left(n\pi \beta\right)\sin\left(n\pi \sqrt{\beta}\bar\delta\right)}\right)\\
&-\frac{1}{4}\sum_{n=1}^\infty\left(\frac{1+\cot\left(\frac{n\pi}{\beta}\right)\cot\left(\frac{n\pi\bar\delta}{\sqrt{\beta}}\right)}{n}+\frac{\cot\left(n\pi\beta\right)\cot\left(n\pi \sqrt{\beta}\bar\delta\right)-1}{n}\right)\,,
\end{split}
}
\eq
where we made use of the periodicity $\cos(x+ n \pi)=\cos(x - n \pi)$. The inversion symmetry \req{betaInversion} is clearly manifest in the non-perturbative part of the free energy for $\beta\not\in \Z$, as expected.  

It remains to write \req{FANPbnZfinal} in terms of $g_s$. Making use of \req{cosSumtoExpSumId} and \req{CotCotIdentity}, we deduce
\beq
\begin{split}
\Fcal_A^{NP}(\bar\mu;\beta\not\in \Q) &\sim -\frac{\ii}{4}\sum_{n=1}^\infty\frac{e^{\frac{-2 n\pi^2 }{\sqrt{\beta} g_s}}Q^{\frac{2\pi \ii n}{\sqrt{\beta} g_s}}}{n\sin\left(\frac{n \pi}{\beta}\right) \sinh\left(\frac{2 n \pi^2 }{\sqrt{\beta}g_s}\right)}-\frac{\ii}{4}\sum_{n=1}^\infty\frac{e^{\frac{-2n\pi^2 \sqrt{\beta}}{g_s}}Q^{\frac{2\pi \ii n \sqrt{\beta}}{ g_s}}}{n\sin\left(n \pi \beta\right)\sinh\left(\frac{2n \pi^2 \sqrt{\beta}}{g_s}\right)}
\\&-\frac{\ii}{2}\sum_{n=1}^\infty\frac{e^{\frac{ \pi n}{\beta g_s}(1+2\pi\ii \sqrt{\beta} g_s) }}{n\sin\left(\frac{n\pi}{\beta}\right)\sinh\left(\frac{2 n\pi^2}{\sqrt{\beta}g_s}\right)}-\frac{\ii}{2}\sum_{n=1}^\infty\frac{e^{\frac{\pi n \beta}{g_s}(1-2\pi\ii g_s/\sqrt{\beta})}}{n\sin\left(n\pi \beta\right)\sinh\left(\frac{2n\pi^2 \sqrt{\beta}}{g_s}\right)}\,.
\end{split}
\eq
(We dropped again some generalized Bernoulli polynomials.)

\paragraph{$\beta\in\Z^*$}
As for our purposes, \ie, comparison with known non-perturbative results at $\beta=1$, it is sufficient to take $\beta\in\Z^*$, we restrict for simplicity here to this case. We have one ray of non-perturbative double poles at $x^*_{np,2}=\pi\ii n$ and one ray of simple poles at $x^*_{np,1}=m\pi \ii/\beta$ with $m\in\N\setminus\{l \beta\}$. We infer
\beq
\begin{split}
R_{np,1}&=-\frac{1}{2}\sum_{n\neq l\beta}^\infty \frac{\sin\left(\frac{n\pi\delta(\mu-1)}{\beta}\right)\sin\left(\frac{n\pi\delta(\mu+1/\delta)}{\beta}\right)}{n\sin\left(\frac{n\pi}{\beta}\right)\sin\left(\frac{n\pi\delta}{\beta}\right)}\\
R_{np,2}&=\sum_{n=1}^\infty\frac{(1-\beta+2\delta)-(1-\beta)\cos\left(2n\pi\delta\right)}{8n\beta\sin^2\left(n\pi\delta\right)}\\
&+\sum_{n=1}^\infty\frac{(\beta-1-2\delta\mu)\cos\left(2n\pi\delta(\mu-1)\right)-(\beta-1+2\delta(1-\mu))\cos\left(2n\pi\delta\mu\right)}{8n\beta\sin^2\left(n\pi\delta\right)}\\
&+\sum_{n=1}^\infty\frac{\sin\left(2n\pi\delta\right)+\sin\left(2n\pi\delta(\mu-1)\right)-\sin\left(2n\pi\delta\mu\right)}{8n^2\pi\beta\sin^2\left(n\pi\delta\right)}\,.
\end{split}
\eq
Note that for $\beta$ integer some of the simple poles are enhanced to double poles and therefore the above residue results are more complicated than before. In order to simplify $R_{np,2}$ further, we redefine $\mu\rightarrow\mu'+1/2$ such that with the help of trigonometric sum-to-product identities we are led to 
\beq
\begin{split}
&R_{np,2}(\mu')=\sum_{n=1}^\infty\left(\frac{1-\beta}{4\beta n}+\frac{\sin\left(2n\pi\delta\right)}{8n^2\pi\beta\sin^2\left(n\pi\delta\right)}+\frac{\delta}{4n\beta\sin^2\left(n\pi\delta\right)}\right)\\
&+\sum_{n=1}^\infty\left(-\frac{\cos\left(2n\pi\delta\mu'\right)}{4n^2\pi\beta\sin\left(n\pi\delta\right)}+\frac{(\beta-1-2\delta \mu')\sin\left(2n\pi\delta\mu'\right)}{4n\beta\sin\left(n\pi\delta\right)}-\frac{\delta\cos\left(2n\pi\delta\mu'\right)\cot\left(n\pi\delta\right)}{4n\beta\sin\left(n\pi\delta\right)}\right)\,.
\end{split}
\eq
Adding up the contributions and with some further trigonometric simplifications,
we conclude that 
\beq\eqlabel{FANPbinZ}
\boxed{
\begin{split}
\Fcal^{NP}_A(\mu';\beta\in\Z^*)\sim&\frac{1}{\beta}\Fcal^{NP}_A(\mu';1)-\frac{1-\beta}{\beta}\sum_{n=1}^\infty\frac{\sin\left(2n\pi\delta\mu'\right)}{4n\sin\left(n\pi\delta\right)}
\\
&+\frac{1}{4}\sum_{n\neq l\beta}^\infty\frac{\cos\left(\frac{n\pi\delta(2\mu'+1/\delta)}{\beta}\right)}{n\sin\left(\frac{n\pi}{\beta}\right)\sin\left(\frac{n\pi\delta}{\beta}\right)}-\frac{1}{4}\sum_{n\neq l\beta}^\infty \frac{\cot\left(\frac{n\pi}{\beta}\right)\cot\left(\frac{n\pi\delta}{\beta}\right)}{n}
\end{split}
}\,,
\eq
with
\beq
\begin{split}
\Fcal^{NP}_A(\mu';1)=&\sum_{n=1}^\infty\left(\frac{\cot\left(n\pi\delta\right)}{4\pi n^2}+\frac{\delta}{4n\sin^2\left(n\pi\delta\right)}\right)\\
&-\sum_{n=1}^\infty\left(\frac{\cos\left(2n\pi\delta\mu'\right)}{4\pi n^2\sin\left(n\pi\delta\right)}+\frac{\delta \mu'\sin\left(2n\pi\delta\mu'\right)}{2n\sin\left(n\pi\delta\right)}+\frac{\delta\cos\left(2n\pi\delta\mu'\right)\cot\left(n\pi\delta\right)}{4n\sin\left(n\pi\delta\right)}\right)\,.
\end{split}
\eq
The $\beta$-dependent non-perturbative completion for $\beta\in \Z^*$ given in \req{FANPbinZ} appears to be non-trivial and constitutes one of the main results of this work. Note that the symmetry \req{betaInversion} is not visible, as it maps $\beta\in\Z^*\rightarrow 1/\beta\in \Q$. The symmetry however dictates the solution for $\Fcal_A(\mu;1/\beta \in \Q)$ with $\beta\in\Z^*$.

Finally, let us inspect $\Fcal_A^{NP}(\mu'; 1)$ in some more detail. The exact resolved conifold free energy at $\beta=1$ has been calculated already previously in \cite{M14}. However, it is interesting to compare the exact result predicted by universal Chern-Simons to other recent predictions. For that, note that under usage of the identities in section \ref{CosSinSumId}, and in terms of $g_s$, we have (up to some generalized Bernoulli polynomials)
\beq\eqlabel{FANPb1gs}
\begin{split}
\Fcal^{NP}_A(\mu';1)&\sim\sum_{n=1}^\infty\frac{1}{4\pi \ii n^2 \sinh\left(\frac{2n\pi^2}{g_s}\right)}\left(1+\frac{2n\pi^2 \, e^{\frac{2n\pi^2 \ii}{g_s} }}{g_s \sinh\left(\frac{2n\pi^2}{g_s}\right)}\right)e^{- \frac{2n\pi^2 \ii}{g_s} } \\
&-\sum_{n=1}^\infty\frac{1}{4\pi\ii n^2\sinh\left(\frac{2n\pi^2}{g_s}\right)}\left(1+\frac{4n\pi^2\ii\mu'}{g_s}-\frac{2n\pi^2 \ii}{g_s}\coth\left(\frac{2n\pi^2}{g_s}\right)\right)e^{-\frac{4n\pi^2\ii\mu'}{g_s}}\,.
\end{split}
\eq
This has to be compared with the prediction for the non-perturbative completion of the topological string on the resolved conifold inferable from the conjecture of \cite{HMMO13}. Under setting 
$$
g_s\rightarrow \frac{4 \pi^2}{\ii\hbar}\,,
$$
we precisely recognize in the second row of \req{FANPb1gs} the conjectured M2-brane contribution, stated for instance in \cite{H15} (\cf, their eq. (5.40)). The first row corresponds to the non-perturbative completion of the constant map contribution, which, to our knowledge, is not given explicitly in the literature. We conclude that the exact calculations via universal Chern-Simons give strong evidence about the validity of the conjecture of \cite{HMMO13} for the resolved conifold. Turned around, this implies that the large $N$ duality is exact. (Very recently, while our work was in the writeup process, some numerical checks of the exactness of the large $N$ duality between Chern-Simons and the unrefined conifold appeared, see \cite{H15b}.)

\section{Example 2: $D_{N/2}$}
\label{Dn}

In this section we will discuss the free energy of refined $D_{N/2}$ Chern-Simons theory. Invoking the multiple sines technique introduced in section \ref{msineSec}, the free energy can be obtained via a simple residue calculation as a Gopakumar-Vafa sum, and non-perturbative corrections thereto, similar as in the previous section.

The parameters leading to $SO(N)$ refined Chern-Simons theory ($N$ is understood to be even) have been derived in \cite{KS13} and read
\beq\eqlabel{DnPara}
v_1=-2\,,\,\,\,\,\, v_2=4\beta\,,\,\,\,\,\, v_3= \beta (N-4)\,,\,\,\,\,\, t=\beta (N-2)\,,\,\,\,\,\, \delta=\kappa+t\,.
\eq
The multiple sine expression of the free energy $\Fcal$ derived in section \ref{msineSec}, \ie, \req{FVviaSine5}, can be used for calculation of free energy in this case. However, without passing through quintuple sine functions, we can infer via a more direct independent calculation that
\beq \eqlabel{pf}
Z_D(N;\beta)=\frac{S_3(1+\frac{3}{2}t|1,t,\delta)}{S_3(1+\delta+\frac{1}{2}t|1,t,\delta)} \times\frac{S_3(1+\beta+t)|1,2\beta,\delta)}{S_3(\beta|1,2\beta,\delta)}\,,
\eq
where $Z_D(N;\beta) := e^{-\Fcal_D}$.
The first factor can be simplified as
$$
\frac{S_3(1+\frac{3}{2}t|1,t,\delta)}{S_3(1+\delta+\frac{1}{2}t|1,t,\delta)} = 
\frac{S_2(1+\frac{1}{2}t|1,t)}{S_2(1+\frac{1}{2}t|1,\delta)}=\frac{S_2(\frac{1}{2}t|1,t)}{S_1(\frac{1}{2}t|t) S_2(1+\frac{1}{2}t|1,\delta)}=\frac{\sqrt{2}}{S_2(1+\frac{1}{2}t|1,\delta)}\,,
$$
where we made use of the identities $S_1(t/2|t)=1$ and $S_2(t/2|1,t)=\sqrt{2}$. Note that the remaining non-trivial denominator $S_2(1+\frac{1}{2}t|1,\delta)$ becomes $S_2(\frac{1}{2}N|1,\delta)$ in the non-refined limit $\beta=1$ and matches one of the multipliers of the partition function of $SO(N)$ Chern-Simons given in \cite{M14} (recall the invariance of multiple sines under simultaneous rescaling of the argument and all parameters, \cf, \req{SrIds}), \ie,  
\begin{eqnarray} \label{SON}
Z_D(N;1)=2^{-\frac{3}{4}}\sqrt{\frac{S_3(2N|2,2,2\delta)}{S_3(2|2,2,2\delta)}} \frac{\sqrt{S_2(2N|4,2\delta)}}{S_2(N|2,2\delta)}\,.
\end{eqnarray}

Let us verify that the other terms in \req{SON} appear from the remaining second factor in the refined partition function \req{pf}. For that, note that the denominator of the second factor in \req{pf} can be rewritten for $\beta=1$ as (see appendix \ref{mgsf})
$$
S_3(1|1,2,\delta)=\sqrt{S_3(1|1,1,\delta)}\sqrt{S_2(1|2,\delta)}=2^{\frac{1}{4}}\sqrt{S_3(1|1,1,\delta)}\,,
$$
which exactly matches the corresponding term in the $SO(N)$ Chern-Simons partition function given above. So, our refined $SO(N)$ partition function equals the $SO(N)$ Chern-Simons partition function at $\beta=1$, as it should be. 

Let us move on to the refined case with general $\beta$. The multiple sine representation \req{pf} of the refined partition function is essentially (we omit numerical multipliers)
\beq\eqlabel{ZDviaS3s}
Z_D(N;\beta)\sim\frac{S_3(1+\beta (N-1)|1,2\beta,\delta)}{S_3(\beta|1,2\beta,\delta)\, S_2(1-\frac{1}{2}\beta+\frac{1}{2} \beta (N-1)|1,\delta)}
\,.
\eq

This form of the partition function is most suited to invoke the integral representation of multiple sine functions \req{SrIntRep} in order to obtain the free energy as a sum of residue (\cf, \req{QSineRes}). However, we do not need the full expression \req{ZDviaS3s} for that. Namely, if we define
$$
\mu:=\frac{\beta N - \beta}{\delta}\,,
$$
and compare with $Z_A$ given in \req{ZaS3final}, we infer that actually
\beq\eqlabel{ZDviaZA}
\boxed{
Z_D(\mu;\beta)\sim Z_A(\mu;2\beta)\times M(\beta)\, e^{\mathcal T(\mu;\beta)}
}\,,
\eq
with
\beq
\begin{split}
\mathcal T(\mu;\beta) &:= \frac{1}{2} \log \mu -\log S_2(1-\beta/2+\delta\mu/2|1,\delta)\,,\\
M(\beta) &:=\frac{S_3(2\beta|1,2\beta,\delta)}{S_3(\beta|1,2\beta,\delta)}\,,
\end{split}
\eq
holds. (We made use of the identities \req{mSineAllIds}). Note that $\mu$ is the usual t'Hooft coupling kept fixed in a large $N$ limit. However, we like to stress that we actually do not take any limit. We observe that the refined $D_{N/2}$ partition function splits into an oriented and unoriented sector, \cf, \cite{KW12}. Usually, $\mathcal T$ is referred to as domainwall tension. The function $M(\beta)$ is the unoriented piece of the constant map contribution and is expected to relate at $\beta=1$ to the real MacMohan function introduced in the context of orientifold in \cite{KSW09,K10}. 

As a side remark, note that the $\Omega$-background symmetry discussed in the $A_{N-1}$ example does not carry over to $D_{N/2}$, as the domainwall tension (or orientifold plane) breaks the symmetry. Instead, it seems that the related transformation 
$$
\beta \rightarrow 1/4\beta\,,\,\,\,\,\,\delta \rightarrow\delta /2 \beta\,,
$$
combined with some appropriate transformation of $N$, maps partition functions of refined $SO(N)$ to refined $Sp(N)$ and vice versa. Since we do not have at hand the refined universal version of the $Sp(N)$ Chern-Simons partition function, we can not explicitly verify this hypothesis at the time being. (However, one may use the expected transformation properties, and the expected structure of the refined partition functions, \cf, \cite{KW12}, to predict the corresponding Vogel parameters.)

As we already discussed the non-perturbative completion of the Gopakumar-Vafa expansion of $Z_A(\mu;\beta)$ in the previous section, we only need to discuss here the additional contribution due to the factors $M$ and $\mathcal T$. Let us start with $\mathcal T$. From the integral representation \req{SrIntRep} we immediately deduce
$$
\mathcal T = \frac{1}{2}\log\mu+ \frac{\ii\pi}{2}B_{2,2}(1-\beta/2+\delta\mu/2|1,\delta)+\int_{\R^+} \frac{dx}{x} \frac{e^{x(1-\beta/2+\delta\mu/2)}}{(e^{x}-1)(e^{\delta x}-1)}\,.
$$ 
(For convenience, we drop from now on the first two terms.) As in the previous $A_N$ discussion, we can distinguish between perturbative and non-perturbative poles (the former depending on $g_s$). Hence, there is a natural split into a perturbative and non-perturbative part, \ie, 
$$
\mathcal T=\mathcal T^P+\mathcal T^{NP}\,.
$$
As we assume $\delta$ not to be integer, there are however no cases to be distinguished. Hence, we can write down immediately
\beq\eqlabel{Tfinal}
\boxed{
\mathcal T(\mu;\beta) \sim -\frac{\ii}{2} \sum_{n=1}^\infty \frac{e^{\ii\pi  n (\mu+(1-\beta)/\delta)}}{n\sin\left(\frac{n \pi}{\delta}\right)}+\frac{\ii}{2}\sum_{n=1}^\infty\frac{e^{n\pi\ii \delta(\mu-1-\beta/\delta)}}{n \sin\left(n \pi \delta\right)}
}\,,
\eq
where the first sum corresponds to $\mathcal T^P$ and the second to $\mathcal T^{NP}$.

For later reference, note that $\mathcal T$ can be as well expressed as a linear combination of the free energies $\Fcal_A$. This can be seen as follows. Using the second identity listed in \req{mSineAllIds} we infer
$$
e^{\mathcal T(\mu;\beta)}=\frac{S_3(1-\beta/2+b+\delta\mu/2|1,b,\delta)}{S_3(1-\beta/2+\delta\mu/2|1,b,\delta)}=\frac{Z_A(\mu/2 +(4b-\beta)/(2\delta);2b)}{Z_A(\mu/2-\beta/(2\delta);2b)}\,, 
$$
where $b\neq 0$ is arbitrary. Hence, taking for instance $b=1/2$, we conclude
\beq\eqlabel{TviaDF}
\mathcal T(\mu;\beta) = \Fcal_A(\mu/2 +(2-\beta)/(2\delta);1)-\Fcal_A(\mu/2-\beta/(2\delta);1)\,.
\eq

It remains to discuss $M(\beta)$. Ignoring for notational convenience the generalized Bernoulli polynomials, we have 
$$
\log M(\beta)\sim \int_{\R^+}\frac{dx}{x}\frac{e^{2\beta x}-e^{\beta x}}{(e^x-1)(e^{2\beta x}-1)(e^{\delta x}-1)}=\int_{\R^+}\frac{dx}{x}\frac{e^{\beta x}}{(e^x-1)(e^{\beta x}+1)(e^{\delta x}-1)}\,.
$$
As is by now familiar, we have a split into a perturbative and non-perturbative piece. Summing the residue of the perturbative poles, yields
$$
\log M^P(\beta)= \frac{\ii}{4}\sum_{n=1}^\infty \frac{e^{\frac{\ii \pi n (\beta-1)}{\delta}}}{n\sin\left(\frac{n\pi}{\delta}\right)\cos\left(\frac{n\pi \beta}{\delta}\right)}\,.
$$
For the non-perturbative parts we have to distinguish between $\beta\in \Q$ or not.

\paragraph{$\beta\not\in \Q$} 
We obtain
$$
\log M^{NP}(\beta\not\in \Q) = \frac{\ii}{4}\sum_{n=1}^\infty \frac{e^{n\pi \ii (\beta-1)}}{n \sin\left(n\pi\beta\right)\cos\left(n\pi \delta\right)}-\frac{1}{2}\sum_{n\,{\rm odd}}\frac{e^{-\frac{\ii\pi n(\delta+1)}{2\beta}}}{n\sin\left(\frac{n\pi}{2\beta}\right)\sin\left(\frac{n\pi \delta}{2\beta}\right)}\,.
$$ 
Note the non-perturbative appearance of an odd sector in $n$. For $\beta\in\Q$ let us just give one example. Namely $\beta=1$ corresponding to (unrefined) topological string orientifolds. 

\paragraph{$\beta = 1$}
We infer
$$
\log M^{NP}(1)=-\frac{i}{2}\sum_{n=1}^\infty (-1)^n \frac{e^{-\frac{\ii n\pi\delta}{2}}}{n \,\sin\left(\frac{n\pi\delta}{2}\right) }\,.
$$
Hence, combining all parts, we deduce via relation \req{ZDviaZA} and under using \req{FANPbinZ} that
\beq
\begin{split}
\Fcal_D^{NP}(\mu';1) \sim &\,\frac{1}{2}\Fcal_A^{NP}(\mu';1)+M^{NP}(1)+\Tcal^{NP}(\mu',1)\\
&+\frac{1}{4}\sum_{n=1}^\infty\frac{\sin\left(2n\pi\delta\,\mu'\right)}{n\sin(n\pi\delta)}-\frac{1}{4}\sum_{n=1}^\infty\frac{\sin(n\pi\delta\mu')}{n\sin\left(\frac{n\pi\delta}{2}\right)}\,,
\end{split}
\eq
where the last two terms are the remaining terms of \req{FANPbinZ} at $\beta=2$ (the fractional sums can be resummed for this value of $\beta$). The first of them, can be eliminated via a further shift of $\mu'$.

\section{Quantum limit}
\label{QL}

It is interesting to consider the quantum limit introduced in \cite{K13} at hand of $\beta$-ensembles. Namely, taking $N$ large with
$$
\Ncal:=\beta N\,,
$$
fixed. In order to keep $\Ncal$ finite, this also requires taking $\beta\rightarrow 0$. Hence, we define
\beq\eqlabel{Wdef}
_0\Wcal := \lim_{\beta\rightarrow 0}\beta\, \Fcal \,.
\eq
(The notation of quantum limit originates from the fact that in this limit $\beta$-ensembles are captured by ordinary quantum mechanics. However, this notation may not be the best choice as it rather corresponds to a classical limit reducing a double quantized system to a single quantized system.)

\paragraph{$A_N$}
Let us apply the quantum limit to the refined $A_N$ case discussed in section \ref{ANsec}. However, it is instrutive to apply the limit not after residue taking, but directly to \req{FintRep}. We infer
$$
_0\Wcal_A(\Ncal) = -\frac{1}{4}\int_{\R^+}\frac{dx}{x^2}\, \frac{\sinh\left(x(\Ncal-\delta)\right)\sinh\left(x(\Ncal+1)\right)}{\sinh\left(x\delta\right)\sinh\left(x\right)}\,.
$$
We have one ray of perturbative simple poles at $x^*_p=n\pi\ii/\delta$ and one ray of non-perturbative simple poles at $x^*_{np}=n\pi\ii$. Taking residue leads to 
$$
_0\Wcal_A(\Ncal) =\, _0\Wcal^P_A(\Ncal) +\, _0\Wcal^{NP}_A(\Ncal)\,,
$$
with
\beq\eqlabel{WAPandNPfinal}
\begin{split}
_0\Wcal_A^P(\Ncal,\delta) &= \frac{\delta}{2\pi}\sum_{n=1}^\infty\frac{\cos\left(\frac{n\pi(2\Ncal+1)}{\delta}\right)}{ n^2 \sin\left(\frac{n\pi}{\delta}\right)}-\frac{\delta}{2\pi}\sum_{n=1}^\infty\frac{\cot\left(\frac{n\pi}{\delta}\right)}{ n^2}\,,\\
_0\Wcal_A^{NP}(\Ncal,\delta) &=\frac{1}{2\pi}\sum_{n=1}^\infty\frac{\cos\left(n\pi(2\Ncal-\delta)\right)}{n^2 \sin\left(n\pi \delta\right)}-\frac{1}{2\pi}\sum_{n=1}^\infty\frac{\cot\left(n\pi \delta\right)}{ n^2}\,.
\end{split}
\eq
(Strictly speaking, the discussion in section \ref{msineSec} about validity of residue taking does not directly apply to the $\beta\rightarrow 0$ limit. However, it is not hard to convince ourselves that the discussion can be adjusted to deal with \req{Wdef}, as in this limit the integrals are actually stronger suppressed.) 

It is interesting to observe that the perturbative and non-perturbative part of $_0\Wcal_A$ are related by
\beq\eqlabel{WANPviaWAP}
\boxed{
_0\Wcal_A^{NP}(\Ncal,\delta)=\frac{1}{\delta}\,_0\Wcal_A^{P}\left(\delta\Ncal-1,\frac{1}{\delta}\right)
}\,,
\eq
We will come back to this S-dual like relation between the perturbative and non-perturbative part of the quantum free energy below. For now, let us take an additional large $\Ncal$ limit, keeping 
\beq\eqlabel{quantNtoMu}
\mu=\frac{\Ncal}{\delta}\,,
\eq
fixed, similar as in \cite{K13}. We infer that 
\beq\eqlabel{WAeqFANS}
\boxed{
_0\Wcal_A(\delta\mu) = \Fcal_{A,NS}(\mu)\,,
}\,.
\eq
where we defined the Nekrasov-Shatashvilli (for short NS) free energy \cite{NS09} (see also \cite{ACDKV11}) as
$$
\Fcal_{A,NS}(\mu):=\lim_{\beta\rightarrow 0} \beta\, \Fcal_A(\mu;\beta)\,,
$$
but there we take now, in contrast to before, the limit after residue taking. The limit can be easily applied to \req{FAPfinal} and \req{FANPbnZfinal}. For that, note that the summation index of the summations in \req{FANPbnZfinal} which can be redefined to run over $n/\beta$ move off to infinity in this limit, and therefore these summations can be dropped. Hence,
\beq
\begin{split}
\Fcal^{P}_{A,NS}(\mu) &= \frac{\delta}{2\pi}\sum_{n=1}^\infty\frac{\cos\left(n\pi(2\mu+1/\delta)\right)}{n^2\sin\left(\frac{n\pi}{\delta}\right)} - \frac{\delta}{2\pi}\sum_{n=1}^\infty\frac{\cot\left(\frac{n\pi}{\delta}\right)}{n^2}\,,\\
\Fcal^{NP}_{A,NS}(\mu) &= \frac{1}{2\pi}\sum_{n=1}^\infty \frac{\cos\left(n\delta(2\mu-1)\right)}{n^2\sin\left(n\pi\delta\right)}-\frac{1}{2\pi}\sum_{n=1}^\infty\frac{\cot\left(n\pi\delta\right)}{ n^2}\,.
\end{split}
\eq
Comparison with \req{WAPandNPfinal} confirms \req{WAeqFANS}. Note however that if we had applied the limit instead to \req{FANPbinZ}, which is only valid for $\beta\in\Z$, we would have gotten a different result. In particular, we learn that $\Fcal_A^{NP}(\mu';1)\neq \Fcal_{A,NS}^{NP}(\mu')$. Instead, it is easy to verify that the relation
\beq\eqlabel{FANPviaFP}
\boxed{
\Fcal_A^{NP}(\mu';1)=\frac{1}{2}\left(\delta \frac{\partial}{\partial \delta} -1\right)\Fcal^{NP}_{A,NS}(\mu')
}\,,
\eq
holds. This relation is an exact statement for the non-perturbative part of refined $A_N$ Chern-Simons on $\S^3$ free energies at large $N$. If we believe in the exactness of the large $N$ duality, this translates to a property of the non-perturbative (refined) topological string free energy on the resolved conifold. Making use of \req{WANPviaWAP}, we can as well express $\Fcal_A^{NP}(\mu';1)$ via $\Fcal^P_{A,NS}$\, leading to the conjecture of \cite{HMMO13} (see also \cite{H15}). 

\paragraph{$D_{N/2}$}
Let us ask what happens in the $D_{N/2}$ case. Applying the quantum limit defined in \req{Wdef} to the integral representation specialized to refined $D_{N/2}$, we observe that 
\beq\eqlabel{0WDviaWA}
_0\Wcal_D(\Ncal) = \frac{1}{2}\, _0\Wcal_A(\Ncal)\,,
\eq
where we kept $\Ncal=\beta N-\beta$ fixed. In fact, this non-uniqueness property of the NS limit has been already observed in \cite{KS11} and \cite{KW12}. Hence, trivially the relation \req{WANPviaWAP} holds as well for $D_{N/2}$. However, this is only one half of the story. As the $\beta$-inversion symmetry is broken in the $D_{N/2}$ case, we have to consider as well the alternative quantum limit
\beq
\eqlabel{InftyWlimit}
_\infty\Wcal := \lim_{\beta\rightarrow \infty} \frac{1}{\beta} \Fcal\,.
\eq
We also define 
$$
\Ncal = N-2\,,
$$
in this limit. (Note that in contrast to the $\beta\rightarrow 0$ case we keep $N$ finite). Physically, after taking $\Ncal$ large with \req{quantNtoMu} fixed, the occurrence of two different limiting cases can be explained by the fact that the orientifold plane lives in two dimensions of the space-time. As the Nekrasov-Shatashvilli limit corresponds to a reduction to two space-time dimensions, we have two choices. Either we reduce to the subspace filled by the orientifold plane, or to the orthogonal subspace. The latter leads to the relation \req{0WDviaWA}, as we do not see the orientifold in space-time anymore. The other case, with orientifold plane, can be investigated as follows.

The limit \req{InftyWlimit} can be calculated via redefining $x\rightarrow x/(2\beta)$, and rescaling $\delta\rightarrow 2\beta\,\delta$ in the integral representation \req{FintRep} (\cf, \cite{KS13}), leading to
$$
_\infty\Wcal_{D} = \frac{1}{2}\int_{\R^+}\frac{dx}{x^2}\left(\frac{\sinh\left(x\,\Ncal/2\right)\sinh\left(x(\Ncal/2-\delta)\right)}{\sinh\left(x\right)\sinh\left(\delta x\right)} +\frac{\sinh\left(x(\Ncal/2-\delta)\right)}{\sinh\left(x\delta\right)}\right)\,.
$$
We denote the integral over the last term as $_\infty \mathcal T=\,_\infty \mathcal T^P+\, _\infty \mathcal T^{NP}$, and evaluate via summation over residue
\beq
\begin{split}
_\infty\mathcal T^P &= \frac{\delta}{\pi}\sum_{n=1}^\infty\frac{\sin\left(\frac{n\pi\Ncal}{2\delta}\right)}{n^2} = \frac{\ii\,\delta}{2\pi} \left({\rm Li}_2\left(e^{-\frac{\ii \pi \Ncal}{2\delta}}\right)-{\rm Li}_2\left(e^{\frac{\ii \pi \Ncal}{2\delta}}\right)\right)\,,\\
_\infty\mathcal T^{NP} &= 0\,.
\end{split}
\eq
Note that $_\infty \mathcal T^{NP}$ vanishes because the second term has only one ray of (perturbative) poles. This is consistent with \req{Tfinal} as in this equation either $\mathcal T^P$ or $\mathcal T^{NP}$ can survive the quantum limit, depending on how we rescale $\delta$ before taking the limit. Evaluating as well the integral over the first term, we conclude 
\beq
\begin{split}
_\infty \Wcal_D^P &= \, - \frac{\delta}{2\pi }\sum_{n=1}^\infty \frac{\cos\left(\frac{n\pi \Ncal}{\delta}\right)}{n^2 \sin\left(\frac{\pi n}{\delta}\right)} + \frac{\delta}{2\pi }\sum_{n=1}^\infty \frac{1}{n^2 \sin\left(\frac{\pi n}{\delta}\right)}+\, _\infty \mathcal T^P\,, \\
_\infty \Wcal_D^{NP}&=\, -\frac{1}{2\pi} \sum_{n=1}^\infty \frac{\cos\left(n\pi (\Ncal-\delta)\right)}{n^2 \sin\left(\pi n \delta \right)}+\frac{1}{2\pi} \sum_{n=1}^\infty \frac{\cot\left(n \pi \delta \right)}{n^2}\,.
\end{split}
\eq
Note that both, the domainwall tension $_\infty\mathcal T$ and the constant map part appear not to be compatible with a relation like \req{WANPviaWAP}.

Finally, let us ask if we nevertheless can still find a relation like \req{FANPviaFP} in the $D_{N/2}$ case. For that, recall from section \ref{Dn} the relation \req{ZDviaZA}. Hence, we have for instance
$$
\Fcal_D^{NP}(\mu;1/2) = \Fcal_A^{NP}(\mu;1)+\log M^{NP}(1/2)+\mathcal T^{NP}\,.
$$
($\mathcal T^{NP}$ does not vanish away from the Nekrasov-Shatashvilli limit, \cf, \req{Tfinal}.) Recall as well that $\mathcal T$ can be expressed as a linear combination of $\Fcal_A(\mu;1)$ with shifted $\mu$ via \req{TviaDF}. (This corresponds to the well-known fact that the orientifold contribution on the conifold is essentially given by half a closed period, see \cite{AAHV02}.) Since \req{FANPviaFP} extends to shifts $\mu\rightarrow \mu+\frac{s}{\delta}$, we deduce that $\Fcal_D^{NP}(\mu;1/2)$ can be expressed via a linear combination of first order linear differential operators in $\delta$ acting on shifted Nekrasov-Shatashvilli free energies $\Fcal_{A,NS}(\mu)$, up to the constant map contribution.

\paragraph{Remarks}
Several remarks are in order. Though the relation \req{WANPviaWAP} looks very appealing, it rather appears to be a special property of the quantum limit of the refined topological string on the resolved conifold (and other genus zero geometries), as such a relation is not obviously visible for the non-perturbatively corrected $\beta$-ensemble \cite{K13} and $\Ncal=2$ supersymmetric gauge theory \cite{K14} (see also \cite{BD15,KT15}) quantum free energies (where the non-perturbative instanton corrections are due to B-cycle instanton tunneling, at least at a particular point in moduli space). In particular, a general validity of the S-dual like relation \req{WANPviaWAP} would trivialize the theory of resurgence. 

However, the quantization condition used in the context of toric Calabi-Yau manifolds in \cite{GHM14}, see in particular \cite{WZH15}, seem to invoke besides the known perturbative part of the quantum periods \cite{ACDKV11} only the S-dual piece of the non-perturbative part of the quantum (or NS) free energies. As this is directly linked with the validity of the conjecture of \cite{HMMO13}, \ie, that the relation \req{FANPviaFP} (combined with \req{WANPviaWAP}) in general provides the non-perturbative completion for topological strings on (general) toric Calabi-Yau manifolds, for which various numerical checks seem to have been performed in the literature, a puzzle arises. Namely, how can this be ?

Possible resolutions could be as follows. Either there are in fact only A-cycle type instanton corrections (\cf, \cite{PS09}) on toric Calabi-Yau manifolds, see however \cite{CESV14}. (Some might see the fact that the poles in the Gopakumar-Vafa expansion arise only at tree-level, \cf, \cite{HMO12}, as evidence.) There is as well a hidden S-dual like relation between the perturbative and non-perturbative parts of the (non-trivial) quantum free energy calculated in \cite{K13,K14} (and so in ordinary quantum mechanics, which would actually be too good to be true). Or, perhaps most appealing, that the dominant instanton action changes over moduli space, \ie, at large volume A-cycle type instantons are dominant, while for instance at conifold points B-cycle instantons are dominant, \cf, \cite{CESV14,K14,BD15}. In the sense that the above simple relations break down away from large volume, due to quantum effects. We believe clarifying these points will be of general interest.

\acknowledgments

The work of D.K. has been supported by the National Research Foundation of
Korea, Grant No. 2012R1A2A2A02046739. The work of R.M. is partially supported by Volkswagen Foundation and  the Science Committee of the Ministry of Science and Education of the Republic of Armenia under contract  13-1C232.

\appendix

\section{Trigonometric sum identities}
\label{TrigSumIdsSec}
\subsection{$cos$ and $sin$ sums}
\label{CosSinSumId}
Consider the summation
$$
\Sigma_+(x|\ul w):=\sum_{n=1}^\infty \frac{\cos\left(2 \pi n x\right)}{n^s \prod_{l=1}^r\sin\left(n w_l\right)}\,.
$$
From the definition of the generalized Bernoulli polynomials \req{GBernoulliDef} we infer the relation
\beq\eqlabel{rSinViaBin}
\frac{1}{\prod_{i=1}^r\sin\left(n w_r \right)} = \sum_{l=0}^\infty (2\ii)^{r} B_{r,l}(|\ul w|/2 |\ul w)  \frac{(\ii n)^{l-r}}{l!}\,.
\eq
Note that in order to arrive at \req{rSinViaBin} we also made use of the identity $B_{r,l}(c z|c \ul w)=c^{l-r}B_{r,l}(z|\ul w)$ (\cf, \cite{NARU03}). Hence, we can rewrite $\Sigma_+$ as
$$
\Sigma_+=\sum_{l=0}^\infty \frac{(2\ii)^r B_{r,l}(|\ul w|/2 |\ul w)(i)^{l-r}}{l!}\sum_{n=1}^\infty n^{l-r-s} \cos(2\pi n x)\,.
$$
Recall the polylogarithm ${\rm Li}_s(z):=\sum_{n=1}^\infty \frac{z^n}{n^s}$ (defined via the convergent series for $|z| <1$, but can be extended to the whole complex plane via analytic continuation), and Jonqui\`ere's inversion formula valid for $n>0$ (and $0\leq \Re x<1$ if $\Im x >0$, else $0 <\Re x\leq 1$)
$$
{\rm Li}_n(e^{2\pi\ii x})+(-1)^n\, {\rm Li}_n(e^{-2\pi\ii x})= -\frac{(2\pi\ii)^n}{n!} B_n(x)\,,
$$
with $B_n(x)$ the ordinary Bernoulli polynomials. The above inversion formula extends to $n<0$ if we define that $B_{n<0}(x)=0$. Hence, in terms of the polylogarithm we can write
\beq
\begin{split}
\Sigma_+=\sum_{l=0}^\infty & \frac{(2\ii)^{r-1} B_{r,l}(|\ul w|/2 |\ul w)(i)^{l-r}}{l!}\\
&\times\left(\left(1-(-1)^{r+s-l}\right) {\rm Li}_{r+s-l}\left(e^{-2\pi\ii x}\right) -\frac{(2\pi\ii)^{r+s-l}}{(r+s-l)!} B_{r+s-l}(x) \right)\,.
\end{split}
\eq
Now note that $B_{r,l}(|\ul w|/2 |\ul w)$ vanishes for $l$ odd due to the identity \req{BshiftId}. But if $l$ is even only, we have
$$
\boxed{
\Sigma_+=\left\{
\begin{matrix}
\Xi(x|\ul w)+\sum_{l=0}^\infty \frac{(2\ii)^r B_{r,l}(|\ul w|/2 |\ul w)(i)^{l-r}}{l!}  \,{\rm Li}_{r+s-l}\left(e^{-2\pi\ii x}\right)&r+s\,{\rm odd}\\
\Xi(x|\ul w)& r+s\,{\rm even}
\end{matrix}
\right.
}\,,
$$
where we defined
$$
\Xi(x|\ul w):=-\sum_{l=0}^\infty  \frac{\ii^{r+s-1}2^{2r+s-l-1}\pi^{r+s-l} B_{r,l}(|\ul w|/2 |\ul w)B_{r+s-l}(x)}{l!(r+s-l)!}\,.
$$
Note that the summation in $\Xi$ is finite (as we defined the Bernoulli polynomials $B_n$ to vanish for negative $n$). Writing the polylogarithm again as a series and using \req{rSinViaBin},
we conclude with the identities valid for $r+s$ odd
\beq\eqlabel{cosSumtoExpSumId}
\boxed{
\sum_{n=1}^\infty \frac{\cos\left(2\pi n x\right)}{n^s \prod_{l=1}^r\sin\left(n w_l\right)}=\left\{
\begin{matrix}
\Xi(x|\ul w)&+\sum_{n=1}^\infty \frac{e^{-2\pi \ii n x}}{n^s \prod_{l=1}^r\sin\left(n w_l\right)}\\
\Xi(1-x|\ul w)&+\sum_{n=1}^\infty \frac{e^{2\pi \ii n x}}{n^s \prod_{l=1}^r\sin\left(n w_l\right)} 
\end{matrix}
\right.  
}\,,
\eq
where the equality in the lower row is due to the fact that we could have equally expressed $\Sigma_+$ in terms of ${\rm Li}_n(e^{+\ii x})$ (making also use of the identity $(-1)^n B_n(x)=B_n(1-x)$).

Similarly, we can consider
$$
\Sigma_-(x|\ul w):=\sum_{n=1}^\infty \frac{\sin\left(2\pi n x\right)}{n^s \prod_{l=1}^r\sin\left(n w_l\right)}\,.
$$
The only difference to $\Sigma_+$ discussed above will be a flip of sign and an additional overall factor of $-\ii$. Hence, we have for $r+s$ even
\beq\eqlabel{sinSumtoExpSumId}
\boxed{
\sum_{n=1}^\infty \frac{\sin\left(2\pi n x\right)}{n^s \prod_{l=1}^r\sin\left(n w_l\right)}=\left\{
\begin{matrix}
-\ii\Xi(x|\ul w)&-\ii\sum_{n=1}^\infty \frac{e^{-2\pi \ii n x}}{n^s \prod_{l=1}^r\sin\left(n w_l\right)}\\
\ii\Xi(1-x|\ul w)&+\ii\sum_{n=1}^\infty \frac{e^{2\pi \ii n x}}{n^s \prod_{l=1}^r\sin\left(n w_l\right)}
\end{matrix}
\right.  
}\,.
\eq
As a side remark, note that for the parity of $r+s$ there the polylogarithms cancel out we provide via the function $\Xi$ a finite expression for some of the summation formula in \cite{FV99}.

\subsection{$\cot$ sum}
\label{cotIdAppendix}
Consider
$$
\Sigma:=\sum_{n=1}^\infty \frac{1+\cot(n\pi x)\cot(n\pi y)}{n}=\sum_{n=1}^\infty\frac{1-\cot\left(n \pi x\right)\cot\left(-n \pi y\right)}{n} \,.
$$
For simplicity, we assume in the following that $\Re y=0$. Invoking the identity 
\beq\eqlabel{cotId}
\pi \cot\pi z = \ii\pi+\frac{2\pi\ii}{e^{2\pi\ii z} -1}\,,
\eq
leads to
$$
\Sigma = \sum_{n=1}^\infty\frac{2}{n} \left(1+\frac{1}{\left(e^{2 n\pi \ii x}-1\right)}-\frac{1}{\left(1-e^{-2 n\pi \ii y}\right)}-\frac{2 e^{-2 n\pi \ii x}}{\left(1-e^{-2 n\pi \ii x}\right)\left(1-e^{-2 n\pi \ii y}\right)}\right)\,.
$$
For $\Im y > 0$ we infer with help of the geometric series
\beq
\begin{split}
\Sigma &= 2\sum_{n=1}^\infty \frac{1}{n\left(e^{2 n\pi \ii x}-1\right)}-2\sum_{n=1}^\infty \frac{1}{n\left(e^{2 n\pi \ii y}-1\right)}+\sum_{n=1}^\infty\frac{ e^{- n\pi \ii (x-y)}}{n \sin\left(n\pi x\right)\sin\left(n\pi y\right)}\\
&=\ii\sum_{n=1}^\infty \frac{\sin\left(n\pi (x-y)\right)}{n \sin\left(n\pi x\right)\sin\left(n\pi y\right)}+\sum_{n=1}^\infty\frac{ e^{- n\pi \ii (x-y)}}{n \sin\left(n\pi x\right)\sin\left(n\pi y\right)}\,,
\end{split}
\eq
where we made use of \req{cotmcotId}. Similarly, for $\Im y < 0$ we have
\beq
\begin{split}
\Sigma&=\ii\sum_{n=1}^\infty \frac{\sin\left(n\pi (x+|y|)\right)}{n \sin\left(n\pi x\right)\sin\left(n\pi |y|\right)}+\sum_{n=1}^\infty\frac{ e^{-n\pi \ii (x+|y|)}}{n\sin\left(n\pi x\right)\sin\left( n\pi |y|\right)}\,.
\end{split}
\eq
Invoking the summation identity \req{sinSumtoExpSumId}, we finally deduce (for $\Re y =0$)
\beq\eqlabel{CotCotIdentity}
\boxed{
\Sigma=\left\{
\begin{matrix}
\,2 \sum_{n=1}^\infty\frac{ e^{- n\pi \ii (x-y)}}{n \sin\left(n\pi x\right)\sin\left(n\pi y\right)}+\Xi(x-y|x,y)& \Im y > 0\\
\,2 \sum_{n=1}^\infty\frac{ e^{- n\pi \ii (x+|y|)}}{n \sin\left(n\pi x\right)\sin\left(n\pi |y|\right)}+\Xi(x+|y||x,|y|)& \Im y < 0
\end{matrix}
\right.
}\,.
\eq

\section{More on multiple gamma and sine functions}
\label{mgsf}

Barnes' multiple zeta functions (as defined in \req{BzetaDef}), fulfill recurrent relations \cite{B1904,R00}, which were implicitly widely used in this paper. Indeed, for $z=z_0+w_i, \Re z_0>0$ the sum over $n_i$  effectively starts from $n_i=1$, with zeta function argument being $z_0$. Adding and subtracting the contribution of $n_i=0$, we get the relation 

\begin{eqnarray}\label{rec}
\zeta_N(z_0+w_i,s|\ul w)=\zeta_N(z_0,s|\ul w)-\zeta_{N-1}(z_0,s|w_1,..., w_{i-1},w_{i+1}...,w_N)\,,
\end{eqnarray}
In turn, from this follows the recurrence relation on multiple gamma-functions:
$$
\Gamma_N(z_0+w_i|w_1,w_2,...,w_N)=\frac{\Gamma_N(z_0|w_1,w_2,...,w_N)}{\Gamma_{N-1}(z_0|w_1,..., w_{i-1},w_{i+1}...,w_N)}\,,
$$

Another type of relations between multiple gamma functions appear when there are some relations between parameters. This is based on the integral representation  \cite{R00} (see \req{GBernoulliDef} for the definition of generalized Bernoulli polynomials)
\beq\eqlabel{Psi}
\begin{split}
\log\Gamma_N(Z|\ul w)=\int_{0}^{\infty}\frac{dx}{x}&\left( e^{-zx} \prod_{j=1}^{N} \frac{1}{(1-e^{-w_jx})} \right.  \\
&\left. - x^  {-N}\sum_{n=0}^{N-1}\frac{(-x)^n}{n!}B_{N,n}(z) -\frac{(-1)^N}{N!}e^{-x}B_{N,N}(z)\right)\,,
\end{split}
\eq
and the observation of \cite{M13} that if in the linear combination of logarithms of some multiple gamma functions the main terms (\ie, the first terms in \req{Psi}) cancel, then all other terms cancel as well. This yields a relation between the corresponding multiple gamma functions. As an example, consider the identity 
$$
\frac{1}{2}\left(\frac{1}{(1-e^{-x})^2}+\frac{1}{1-e^{-2x}}\right)=\frac{1}{(1-e^{-x})(1-e^{-2x})}\,,
$$
which implies identities between multiple gamma functions like
\beq\label{qwer}
\begin{split}
\Gamma_2(N|1,2)&=\sqrt{\Gamma_2(N|1,1)\Gamma_1(N|2)}\,,\\
\Gamma_3(N|1,2,x)&=\sqrt{\Gamma_3(N|1,1,x)\Gamma_2(N|2,x)}\,,
\end{split}
\eq
\etc \,. 

Evidently, such relations can appear each time one has rational relations between parameters. We also note that from the identity \cite{R00}
$$
\Gamma_1(w|a)=\exp\left( \left( \frac{w}{a}-\frac{1}{2}\right) \ln a \right) \Gamma\left( \frac{w}{a}\right) (2\pi)^{-\frac{1}{2}}\,,
$$
it follows
$$
\Gamma_1(x|x)=\sqrt{\frac{x}{2\pi}}\,,\,\,\,\,\,\Gamma_1(x|2x)=\sqrt{\frac{1}{2}}\,,\,\,\,\,\,\Gamma_0(w)=1/w\,.
$$
where the last identity holds by definition.

Recalling the definition of the multiple sine functions in terms of gamma functions \req{mSineGdef}, it is now easy to show that we have
\beq
\begin{split}
S_3(N|1,2,\delta)&=\frac{1}{\Gamma_3(N|1,2,\delta) \Gamma_3(\delta+3-N|1,2,\delta)}=\frac{\Gamma_2(\delta+2-N|2,\delta)}{\Gamma_3(N|1,2,\delta) \Gamma_3(\delta+2-N|1,2,\delta)}\\
&=\frac{\Gamma_2(\delta+2-N|2,\delta)}{\sqrt{\Gamma_3(N|1,2,\delta)\Gamma_2(N|2,\delta)} \sqrt{\Gamma_3(\delta+2-N|1,2,\delta)\Gamma_2(\delta+2-N|2,\delta)}}\\
&=\sqrt{S_3(N|1,1,\delta)}\sqrt{S_2(N|2,\delta)}\,.
\end{split}
\eq
This identity is made use of in section \ref{Dn}.

Clearly, as similar integral representations as \req{Psi} exists for the multiple sine functions (\cf, \req{SrIntRep}) \cite{NARU03}, one can as well derive new identities for multiple sine functions in a similar spirit. For instance, from the relation
\beq\eqlabel{exprel}
\frac{1}{2}\left(\frac{1}{(e^{x}-1)^2}+\frac{1}{e^{2x}-1}\right)=\frac{e^x}{(e^{x}-1)(e^{2x}-1)}\,,
\eq
we can deduce the identity
\beq
\sqrt{\frac{S_3(N|1,1,y)}{S_2(N|2,y)}}=S_3(N+1|1,2,y)\,.
\eq
As a side remark, note that for the multiple sines it is even more evident that relations of the type \req{exprel} lead to identities between multiple sine functions, since both terms in the integral representation, \ie, integral over real line and generalized Bernoulli polynomial (the residue contribution of the origin), are integrals of the same integrand of type \req{exprel}.

Finally, for the readers convenience, we list some of the simpler identities of sine functions we made heavily use of in the main text
\beq\eqlabel{mSineAllIds}
\begin{split}
S_r(cz|c\underline{\omega})&=S_r(z|\underline{\omega})\,,\\
S_r(z+\omega_i|\underline{\omega})&=S_r(z|\underline{\omega})/S_{r-1}(z|\underline{\omega}_i^-)\,,\\
 S_r(z|\underline{\omega})&=S_r(|\underline{\omega}|-z|\underline{\omega})^{(-1)^{r+1}}\,,\\
 S_1(z|\omega)&=2 \sin\frac{\pi z}{\omega}\,, \\
 S_2(1|1,x)&=\sqrt x\,,
\end{split}
\eq
were $\; \underline{\omega}_i^- :=(\omega_1,...,\omega_{i-1},\omega_{i+1},...\omega_r)$.


\begin{thebibliography}{99}
\addcontentsline{toc}{section}{References}
\renewcommand{\itemsep}{-.2cm}
\colorlinksblue
\small

\bibitem{GV98}
  R.~Gopakumar and C.~Vafa,
  ``On the gauge theory / geometry correspondence,''
  Adv.\ Theor.\ Math.\ Phys.\  {\bf 3} (1999) 1415
  [\hepth{9811131}].

\bibitem{SV00}
  S.~Sinha and C.~Vafa,
  ``SO and Sp Chern-Simons at large N,''
  \hepth{0012136}.

\bibitem{N02}
  N.~A.~Nekrasov,
  ``Seiberg-Witten prepotential from instanton counting,''
  Adv.\ Theor.\ Math.\ Phys.\  {\bf 7} (2004) 831
  \hepth{0206161}.

\bibitem{HIV03}
  T.~J.~Hollowood, A.~Iqbal and C.~Vafa,
  ``Matrix models, geometric engineering and elliptic genera,''
  JHEP {\bf 0803} (2008) 069
  [\hepth{0310272}].

\bibitem{IKV07}
  A.~Iqbal, C.~Kozcaz and C.~Vafa,
  ``The Refined topological vertex,''
  JHEP {\bf 0910} (2009) 069
  [\hepth{0701156}].

\bibitem{KW14}
  D.~Krefl and J.~Walcher,
  ``B-Model Approaches to Instanton Counting,''
  \arxiv{1412.7133}{hep-th}.

\bibitem{AS11}
  M.~Aganagic and S.~Shakirov,
  ``Knot Homology and Refined Chern-Simons Index,''
  Commun.\ Math.\ Phys.\  {\bf 333} (2015) 1,  187
  [\arxiv{1105.5117}{hep-th}].

\bibitem{AS12a}
  M.~Aganagic and S.~Shakirov,
  ``Refined Chern-Simons Theory and Knot Homology,''
  Proc.\ Symp.\ Pure Math.\  {\bf 85} (2012) 3
  [\arxiv{1202.2489}{hep-th}].

\bibitem{AS12}
  M.~Aganagic and K.~Schaeffer,
  ``Orientifolds and the Refined Topological String,''
  JHEP {\bf 1209} (2012) 084
  [\arxiv{1202.4456}{hep-th}].


\bibitem{MV12}
  R.~L.~Mkrtchyan and A.~P.~Veselov,
  ``Universality in Chern-Simons theory,''
  JHEP {\bf 1208} (2012) 153
  [\arxiv{1203.0766}{hep-th}].

\bibitem{M13}
  R.~L.~Mkrtchyan,
  ``Nonperturbative universal Chern-Simons theory,''
  JHEP {\bf 1309} (2013) 054
  [\arxiv{1302.1507}{hep-th}].

\bibitem{M13b}
  R.~L.~Mkrtchyan,
  ``Universal Chern-Simons partition functions as quadruple Barnes` gamma-functions,''
  JHEP {\bf 1310} (2013) 190
  [\arxiv{1309.2450}{hep-th}].

\bibitem{KS13}
  D.~Krefl and A.~Schwarz,
  ``Refined Chern-Simons versus Vogel universality,''
  J.\ Geom.\ Phys.\  {\bf 74} (2013) 119
  [arXiv:1304.7873 [hep-th]].

\bibitem{V99}
P.~Vogel {\it Algebraic structures on modules
of diagrams}, preprint (1995), J. Pure Appl. Algebra {\bf 215} (2011), no.
6, 1292-1339.


\bibitem{M81}
R.L.~Mkrtchyan, "The equivalence of Sp(2n) and SO(-2n) gauge theories", Phys.Lett. 105B (1981), 174. 

\bibitem{M14}
  R.~L.~Mkrtchyan,
  ``On a Gopakumar-Vafa form of partition function of Chern-Simons theory on classical and exceptional lines,''
  JHEP {\bf 1412} (2014) 171
  [\arxiv{1410.0376}{hep-th}].

\bibitem{HMMO13}
  Y.~Hatsuda, M.~Marino, S.~Moriyama and K.~Okuyama,
  ``Non-perturbative effects and the refined topological string,''
  JHEP {\bf 1409} (2014) 168
  [\arxiv{1306.1734}{hep-th}].

\bibitem{H15}
  Y.~Hatsuda,
  ``Spectral zeta function and non-perturbative effects in ABJM Fermi-gas,''
  \arxiv{1503.07883}{hep-th}.

\bibitem{K13}
  D.~Krefl,
  ``Non-Perturbative Quantum Geometry,''
  JHEP {\bf 1402} (2014) 084
  [\arxiv{1311.0584}{hep-th}].

\bibitem{W10}
  E.~Witten,
  ``Analytic Continuation Of Chern-Simons Theory,''
  \arxiv{1001.2933}{hep-th}.

\bibitem{B1904}
E.W. Barnes, 
``On the theory of the multiple gamma function,'' 
Trans. Cambridge Philos. Soc.
19 (1904), 374-425.

\bibitem{R00}
S. N. M. Ruijsenaars, 
``On Barnes' Multiple Zeta and Gamma Functions, '' 
Advances in Mathematics 156, 107-132 (2000).

\bibitem{NARU03}
Atsushi Narukawa, {\it The modular properties and the integral representations of the multiple elliptic gamma functions}, Adv. in Math. 189 (2) (2004) 247-267, \math{0306164} [math.QA]

\bibitem{F12}
L. D. Faddeev, {\it Volkov's Pentagon for the Modular Quantum Dilogarithm},
Functional Analysis and Applications, vol.45(4), 2011, p.65, 
\arxiv{1201.6464}{math.QA}. 


\bibitem{FV99}
G.~Felder, A.~Varchenko,
``The elliptic Gamma function and $\rm{SL}(3,\Z)\ltimes \Z^3$''
\math{9907061} [math.QA]

\bibitem{DV09}
  R.~Dijkgraaf and C.~Vafa,
  ``Toda Theories, Matrix Models, Topological Strings, and N=2 Gauge Systems,''
  \arxiv{0909.2453}{hep-th}.

\bibitem{KW10a}
  D.~Krefl and J.~Walcher,
  ``Extended Holomorphic Anomaly in Gauge Theory,''
  Lett.\ Math.\ Phys.\  {\bf 95} (2011) 67
  [\arxiv{1007.0263}{hep-th}].

\bibitem{KW10b}
  D.~Krefl and J.~Walcher,
  ``Shift versus Extension in Refined Partition Functions,''
  \arxiv{1010.2635}{hep-th}.

\bibitem{H15b}
  Y.~Hatsuda and K.~Okuyama,
  ``Resummations and Non-Perturbative Corrections,''
  \arxiv{1505.07460}{hep-th}.

\bibitem{KW12}
  D.~Krefl and J.~Walcher,
  ``ABCD of Beta Ensembles and Topological Strings,''
  JHEP {\bf 1211} (2012) 111
  [\arxiv{1207.1438}{hep-th}].

\bibitem{KSW09}
  D.~Krefl, S.~Pasquetti and J.~Walcher,
  ``The Real Topological Vertex at Work,''
  Nucl.\ Phys.\ B {\bf 833} (2010) 153
  [\arxiv{0909.1324}{hep-th}].

\bibitem{K10}
  D.~Krefl,
  ``Wall Crossing Phenomenology of Orientifolds,''
  \arxiv{1001.5031}{hep-th}.

\bibitem{NS09}
  N.~A.~Nekrasov and S.~L.~Shatashvili,
  ``Quantization of Integrable Systems and Four Dimensional Gauge Theories,''
  \arxiv{0908.4052}{hep-th}.

\bibitem{ACDKV11}
  M.~Aganagic, M.~C.~N.~Cheng, R.~Dijkgraaf, D.~Krefl and C.~Vafa,
  ``Quantum Geometry of Refined Topological Strings,''
  JHEP {\bf 1211} (2012) 019
  [\arxiv{1105.0630}{hep-th}].

\bibitem{KS11}
  D.~Krefl and S.~Y.~D.~Shih,
  ``Holomorphic Anomaly in Gauge Theory on ALE space,''
  Lett.\ Math.\ Phys.\  {\bf 103} (2013) 817
  [\arxiv{1112.2718}{hep-th}].

\bibitem{AAHV02}
  B.~S.~Acharya, M.~Aganagic, K.~Hori and C.~Vafa,
  ``Orientifolds, mirror symmetry and superpotentials,''
  \hepth{0202208}.

\bibitem{K14}
  D.~Krefl,
  ``Non-Perturbative Quantum Geometry II,''
  JHEP {\bf 1412} (2014) 118
  [\arxiv{1410.7116}{hep-th}].

\bibitem{BD15}
  G.~Başar and G.~V.~Dunne,
  ``Resurgence and the Nekrasov-Shatashvili limit: connecting weak and strong coupling in the Mathieu and Lamé systems,''
  JHEP {\bf 1502} (2015) 160
  [\arxiv{1501.05671}{hep-th}].

\bibitem{KT15}
  A.~K.~Kashani-Poor and J.~Troost,
  ``Pure N=2 Super Yang-Mills and Exact WKB,''
  \arxiv{1504.08324}{hep-th}.

\bibitem{GHM14}
  A.~Grassi, Y.~Hatsuda and M.~Marino,
  ``Topological Strings from Quantum Mechanics,''
  \arxiv{1410.3382}{hep-th}.

\bibitem{WZH15}
  X.~Wang, G.~Zhang and M.~x.~Huang,
  ``A New Exact Quantization Condition for Toric Calabi-Yau Geometries,''
  \arxiv{1505.05360}{hep-th}.

\bibitem{PS09}
  S.~Pasquetti and R.~Schiappa,
  ``Borel and Stokes Nonperturbative Phenomena in Topological String Theory and c=1 Matrix Models,''
  Annales Henri Poincare {\bf 11} (2010) 351
  [\arxiv{0907.4082}{hep-th}].

\bibitem{CESV14}
  R.~Couso-Santamaría, J.~D.~Edelstein, R.~Schiappa and M.~Vonk,
  ``Resurgent Transseries and the Holomorphic Anomaly: Nonperturbative Closed Strings in Local ${\mathbb{C}\mathbb{P}^2}$,''
  Commun.\ Math.\ Phys.\  {\bf 338} (2015) 1,  285
  [arXiv:1407.4821 [hep-th]].

\bibitem{HMO12}
  Y.~Hatsuda, S.~Moriyama and K.~Okuyama,
  ``Instanton Effects in ABJM Theory from Fermi Gas Approach,''
  JHEP {\bf 1301} (2013) 158
  [\arxiv{1211.1251}{hep-th}].



\end{thebibliography}
\end{document}